\documentclass[9pt,english,sigconf]{acmart}
\usepackage[T1]{fontenc}
\usepackage[latin9]{inputenc}
\usepackage{babel}
\usepackage{array}
\usepackage{enumitem}
\usepackage{varwidth}
\usepackage{graphicx}
\usepackage{rotating}
\ifx\hypersetup\undefined
  \AtBeginDocument{%
    \hypersetup{unicode=true,pdfusetitle,
 bookmarks=true,bookmarksnumbered=false,bookmarksopen=false,
 breaklinks=true,pdfborder={0 0 0},pdfborderstyle={},backref=false,colorlinks=false}
  }
\else
  \hypersetup{unicode=true,pdfusetitle,
 bookmarks=true,bookmarksnumbered=false,bookmarksopen=false,
 breaklinks=true,pdfborder={0 0 0},pdfborderstyle={},backref=false,colorlinks=false}
\fi

\makeatletter

\acmConference[IPSN]{Information Processing in Sensor Networks}{May 2022}{Milan, Italy}{}

\providecommand{\tabularnewline}{\\}

\usepackage{blindtext,tikz}
\usetikzlibrary{calc}

 \usepackage{url}
\renewcommand\footnotetextcopyrightpermission[1]{}

\makeatother

\begin{document}

\title{Vildehaye: A Family of Versatile, Widely-Applicable, and Field-Proven
Lightweight Wildlife Tracking and Sensing Tags}
\author{Sivan Toledo}
\author{Shai Mendel}
\affiliation{Tel Aviv University}
\author{Anat Levi}
\author{Yoni Vortman}
\affiliation{Tel Hai College}
\author{Wiebke Ullmann}
\author{Lena-Rosa Scherer}
\author{Jan Pufelski}
\affiliation{University of Potsdam}
\author{Frank van Maarseveen}
\author{Bas Denissen}
\author{Allert Bijleveld}
\affiliation{Royal Netherlands Institute for Sea Research}
\author{Yotam Orchan, Yoav Bartan}
\author{Sivan Margalit}
\author{Idan Talmon}
\author{Ran Nathan}
\affiliation{The Hebrew University of Jerusalem}
\begin{abstract}
We describe the design and implementation of \emph{Vildehaye}, a family
of versatile, widely-applicable, and field-proven tags for wildlife
sensing and radio tracking. The family includes 6 distinct hardware
designs for tags, 3 add-on boards, a programming adapter, and base
stations; modular firmware for tags and base stations (both standalone
low-power embedded base stations and base stations tethered to a computer
running Linux or Windows); and desktop software for programming and
configuring tags, monitoring tags, and downloading and processing
sensor data. The tags are versatile: they support multiple packet
formats, data rates, and frequency bands; they can be configured for
minimum mass (down to less than 1~g), making them applicable to a
wide range of flying and terrestrial animals, or for inclusion of
important sensors and large memories; they can transmit packets compatible
with time-of-arrival transmitter-localization systems, tag identification
and state packets, and they can reliably upload sensor data through
their radio link. The system has been designed, upgraded, and maintained
as an academic research project, but it has been extensively used
by 5 different groups of ecologists in 4 countries over a period of
5 years. More than 7100 tags have been produced and most of these
have been deployed. Production used 41 manufacturing runs. The tags
have been used in studies that so far resulted in 9 scientific publications
in ecology (including in \emph{Science}). The paper describes innovative
design aspects of Vildehaye, field-use experiences, and lessons from
the design, implementation, and maintenance of the system. Both the
hardware and software of the system are open.
\end{abstract}
\maketitle

\noindent
\tikz[overlay,remember picture]
{
    \node at ($(current page.north)-(0,1cm)$) 
      [text width=100mm]
      {\tt{Accepted to the ACM/IEEE International Conference on Information Processing in Sensor Networks (IPSN), 
           May 2022,
           Copyright IEEE.}};
}

\section{Introduction}

\renewcommand{\shortauthors}{Toledo et al.}Despite rapid and ongoing
advances in tracking, sensing, and communication technologies, studying
the movement, behavior, and physiology of many species of wild animals
remains highly challenging~\citep{ScienceReview}. Most bird and
bat species, as well as many small mammals and reptiles, can only
carry miniature tracking tags with short antennas that must not entangle.
Solar energy harvesting is not an option for nocturnal and underground
species (including most bats and rodents). Consequently, technologies
designed for humans and their belongings are often poorly matched
to animal sensing and tracking.

This paper presents \emph{Vildehaye} (VH for short) a family of versatile,
widely-applicable, and field-proven lightweight wildlife tracking
tags, base stations that communicate with them, and software that
processes data gathered by tags. VH tags are tiny, down to less than
1~g, making them applicable to a wide range of species. They can
be used for regional high-throughput radio tracking using the ATLAS
system~\citep{ChristineValidationArxiv,ATLAS-PRE-PROCESSING,atlas-accuracy}.
VH tags are modular: add-on boards with sensors and large non-volatile
memories can be attached to the basic radio tag, allowing tags to
sense the behavior and environment of the animal and to log the data.
The data is retrieved either by recapturing the animal or by uploading
data via radio to a nearby base station. The tags can also upload
via radio data from existing specialized wildlife data loggers. A
VH tag can turn on an actuator upon reception of a radio command;
this is used to release data loggers attached to wild birds. The
tags are available for two popular license-free UHF frequency bands
and can be easily modified to support many other VHF and UHF bands.
VH tags come in several variants featuring different tradeoffs between
size and functionality.

VH tags are easy for ecologists to use. A spreadsheet allows researchers
to predict the lifespan of a particular variant with a particular
battery under a given radio schedule. Printed circuit boards (PCBs)
can be ordered directly from a manufacturer in any quantity. Tags
are programmed and configured using a simple and inexpensive FTDI
USB-to-serial dongle and dedicated software. Users routinely provide
and share know-how on how to attach batteries and antennas, how to
weatherproof tags, and how to attach them. 

Consequently, at least 7100 VH tags have been deployed to successfully
track and sense a wide range of wild animals. Users span 5 separate
ecology research groups in Israel, the Netherlands, the UK, and Germany,
and  additional groups in additional counties are in the process
of adopting the tags. Data collected using VH tags have already been
analyzed and published in several research articles in the scientific
(ecology) literature, including \emph{Science}~\citep{EcologyLettersHabitatSelection,SpatialCognitiveAbilityATLAS,OwlMicrobial,PheasantCognitionATLAS,ATLAS-BAT-SPATIAL-PARTITIONING,Toledo188,vilk2021ergodicity},
and additional articles are in preparation.

This paper enumerates requirements that users presented the design
team (Section~\ref{sec:Requirements}) and it presents the resulting
design from the users' viewpoint (Section~\ref{sec:User-Facing-Design}).
Section~\ref{sec:Technical-Design} presents the detailed design
of the tags and their associated software. The discussion emphasizes
innovative aspects related to energy efficiency and effective use
of miniature batteries, to the radio protocol, and to effective logging
of sensor data to flash. Section~\ref{sec:Not-Implemented} discusses
certain features that we decided not to implement. Section~\ref{sec:Use-Cases}
describes four distinct use cases of the tags, three of which have
been used in the field, to demonstrate the versatility and robustness
of the design. We conclude the paper with a discussion of related
work in Section~\ref{sec:Related-Work} and with our conclusions
from the project in Section~\ref{sec:Lessons-Learned}.

\section{\label{sec:Requirements}Requirements }

The requirements for VH tags were largely defined by users. The initial
motivation came from user feedback on the first generation of tags
designed for the first ATLAS tracking system~\citep{TagsEDERC2014}.
These users expressed a set of needs and desires that led to the initiation
of the Vildehaye project in the spring of 2016. As the project progressed,
additional requirements came up as users imagined new applications
for the tags.

Requirements whose importance was ranked high by users include:

\begin{itemize}[leftmargin=*]
\item Compatibility with ATLAS; tags must be able to transmit unique pseudo-random
data packets at high data rates at an accurate ping-repetition interval~\citep{atlas-accuracy}.
\item Mass that is as low as possible, ideally down to 1~g or less, for
wide applicability; ecologists typically limit the mass of a tag to
around 3\% of the mass of the animal, and many species of interest
are small (and thus challenging to track and sense). The effects of
tracking and sensing tags have been extensively investigated; in general,
tags cause some detrimental effects~\citep{meta-analysis-of-biologging-effects,animal-tags-engineering-perspective},
but their use is nonetheless ethically and scientifically justified
in many cases. Smaller tags have smaller effects, so they are preferable.
\item On-board sensors, especially accelerometers and altimeters, to provide
additional information on the location, environment, and behavior
of the animal.
\item Modular hardware and/or multiple hardware variants, to allow production
of both very lightweight tags with limited functionality and lifespan,
and heavier tags with sensors, large memories, and long lifespans;
in particular, multiple battery configurations must be supported.
\item Upload of sensor measurements from tags to base stations via a radio
link, to avoid the need to retrieve the tag in order to collect the
data. Users understand that radio data transfer uses battery energy
and shortens the lifespan of tags powered by primary batteries.
\item Easy programming and configuration (ping-repetition intervals, switching
between intervals, sensing schedules, etc.).
\item Low-cost manufacturing in both small and large batches.
\end{itemize}

Over time, users came up with a few additional requests. The requests
were examined; some were addressed but not all:

\begin{itemize}[leftmargin=*]
\item Transmitting tag-identification packets that can be received by a
base station (ideally a low-power base station that can be powered
by a primary battery) to indicate proximity of the tag to the base
station at distances of up to a few or even tens of kilometers; accepted
and implemented at ranges of up to a few kilometers.
\item The ability to transmit a command to a tag to either change its operational
schedule (e.g., change ping-repetition interval, attempt data upload)
and/or to activate an actuator; implemented.
\item The ability to detect close-range encounters between tags even outside
the range of an ATLAS tracking system.
\item The ability to interoperate with other wildlife radio tracking and
sensing systems; implemented with respect to one such system; there
are plans to expand.
\end{itemize}

At a high level, the design of VH tags aims to be as \emph{widely-applicable}
and as \emph{versatile} as possible in the sense that it should address
as many use cases that require \emph{lightweight radio tags }as possible.
The design does not aim to address use cases that can tolerate high
weights, high power consumption, or completely different tracking
modalities~\citep{ScienceReview}, such as GNSS localization~\citep{Science:GPS-tags},
underwater ultrasound localization~\citep{InjectibleUltrasonicTransmitter},
underground magneto-inductive localization~\citep{HiddenLivesUndergrandAnimals,MEE:Magneto-Inductive},
or data upload through cellular networks; these use cases require
different hardware and firmware architectures so trying to support
them would defocus the project.

\section{\label{sec:User-Facing-Design}User-Facing Design}

\noindent 
\begin{figure}
\begin{centering}
\includegraphics[scale=0.75]{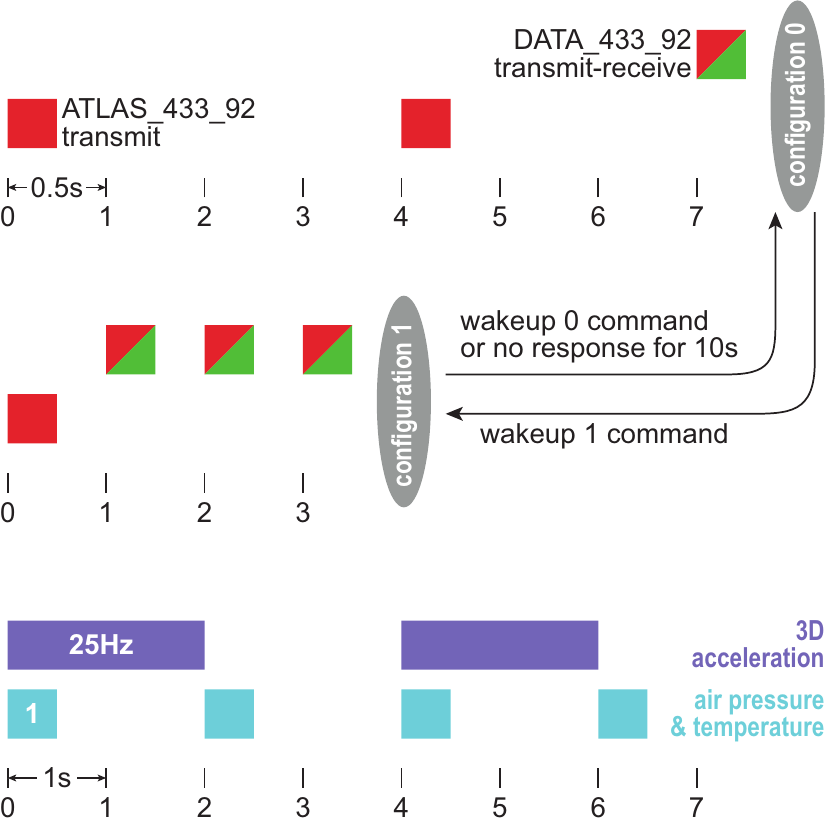}
\par\end{centering}
\caption{\label{fig:tag-def}A sample definition of the behavior of a tag.
The first two rows show two radio configurations, 0 and 1, each with
two radio setups called ATLAS\_433\_92 and DATA\_433\_92. Configurations
are states in a finite-state machine; arrows indicate transition rules.
Red squares represent transmit-only slots. Red/green squares represent
a slot in which the tag transmits and then listens for a reply from
a base station. Acceleration is sensed in bursts, at 25~Hz for 2
seconds every 4~s; air pressure and temperature are sensed once every
2~s. The width of colored squares is not to scale.}

\end{figure}

\noindent \textbf{\emph{Defining Tag Behavior}}. Users define the
radio behavior and sensor behavior of a tag using abstractions shown
graphically in Figure~\ref{fig:tag-def}. The radio behavior is defined
in terms of a user-specified \emph{period}, here 0.5~s, a set of
predefined \emph{radio setups}, which define the frequency, modulation,
symbol rate, packet format, etc.~(here ATLAS\_433\_92 and DATA\_433\_92),
and a set of \emph{configurations}. Each configuration defines a schedule
for the radio in terms of a fixed number of \emph{slots} spaced one
period apart. Slots are allocated to radio setups using a cyclic allocation
with a given starting slot. In our example, configuration~1 repeats
every 8 slots; ATLAS\_433\_92 is allocated every 4th slot starting
at slot~0; DATA\_433\_92 is allocated every 8th slot, starting at~7.
The slots of the ATLAS setup are transmit-only slots. The slots of
the DATA setup are marked as slots that transmit and then listen for
a reply from a base station. The definition of the tag also specifies
the state machine that controls transitions between configurations.
Here, transition from configuration~0, defined as the initial configuration,
to configuration~1 occurs when the tag receives a \emph{wakeup} command
with argument~1 from a base station. The transition back occurs either
when a wakeup~0 command is received, or when the tag has not heard
any reply from a base station for 10~s. The actual transmission and
reception periods are short, typically a few milliseconds long.

The sensing schedule is defined using a fixed 1~Hz grid. For each
available sensor, the user defines how often it is sampled and whether
the samples are one-shot or repetitive. For repetitive sampling the
user specifies the length of the sampling period and the sampling
rate. The user can also specify the configuration of each sensor (what
quantities a multimodal sensor should sense, full-scale, etc). The
definition in Figure~\ref{fig:tag-def} specifies sampling air pressure
and temperature once every 2~s and acceleration every 4~s at in
bursts of 25~Hz for 2~s.

Users can define the behavior of specific tags as well as general
templates applied to tags. Tags have unique individual identification
numbers even when defined from templates.  All the definitions are
stored in version-controlled text files.

\noindent 
\begin{figure}
\begin{centering}
\includegraphics[width=1\columnwidth]{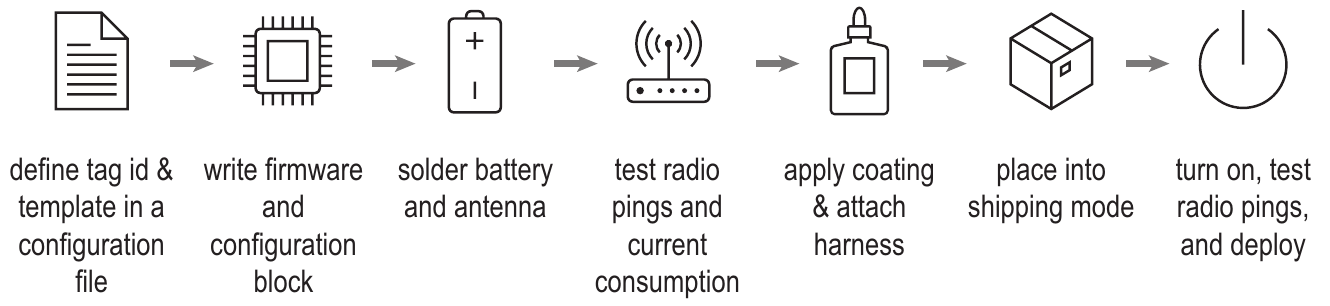}
\par\end{centering}
\caption{\label{fig:tag-preparation}The steps of preparing a VH tag.}

\end{figure}

\noindent \textbf{\emph{Preparing Tags for Deployment}}. Figure~\ref{fig:tag-preparation}
summarizes the process of preparing a VH tag. The user assigns a unique
ID to the tag and defines its behavior, usually using an existing
template. The user attaches a blank assembled tag PCB to an inexpensive
USB-to-serial dongle and invokes software on a Windows computer to
programs the appropriate firmware onto the tag, as well as a block
of binary data that defines the behavior of the tag. Tags can be programmed
over and over again. Physical preparation starts with soldering an
antenna and a primary battery (single Lithium cell or 2 Silver Oxide
cells) to the tag and testing that it transmits. Testing is done either
using a VH base station or using a low-cost software-defined radio
receiver (SDR) capable of detecting ATLAS transmissions. Next, the
tag is coated with insulating varnish~\citep{3M:1601} and optionally
protected from physical damage (biting, scratching) using epoxy (\citep{Kukdo:KH-816,Kukdo:YD-114EF},
sometimes mixed with glass bubbles to reduce weight, or~\citep{3M:DP270});
for other options for coating and protecting tags, see~\citep{InjectibleUltrasonicTransmitter}.
If the tag is deployed (attached to an animal) soon after production,
it is often left active between production and deployment. Otherwise,
the tag is put into a \emph{shipping mode} in which it consumes little
or no power; see below for the mechanisms.. When an add-on memory
board is attached to a tag, the memory board must be \emph{formatted
}prior to attaching to the tag. Formatting the memory board is done
using the same USB-to-serial dongle. The two PCBs are then mated using
stacking board-to-board connectors, usually glued, and are coated
together.

\noindent \textbf{\emph{Collecting Data from Tags}}. Deployed tags
can be tracked by ATLAS and VH base stations (ATLAS produces accurate
localizations, VH base stations only a record of detecting the tag).
Data collected by onboard sensors is retrieved in one of two ways.
The tag can upload the data to a VH base station via a radio link,
or the tag is physically retrieved and the data is downloaded either
using the USB-to-serial dongle or by attachment of the memory board
to the SPI bus of a Rasberry Pi (this is faster). Physical retrieval
can be achieved either by recapturing the animal, or by locating and
collecting a tag that was detached from the animal. Detachment can
be achieved with both passive release mechanisms (an attachment mechanism
that disintegrates over time) or using an active mechanism that is
turned on by sending a command from a base station.

\noindent \textbf{\emph{Processing Collected Sensor Data}}. Sensor
data is stored on tags using a data structure called a\emph{ log},
which represents a sequence of pieces of data called \emph{log items}.
Data collected from tags is transferred an SQL database in raw form,
each row storing a log item as a binary blob. Software running on
Linux or Windows reads the log items associated with a particular
tag and produces binary numeric files containing time-stamped sensor
measurements. These files are read by Matlab or R software that performs
further processing, such as converting air-pressure data to altitude,
classifying behavior based on accelerometer data, and so on. Similar
Matlab and R software is used to process location information from
ATLAS tracking~\citep{ATLAS-PRE-PROCESSING}. This analysis software
is not part of the VH system and will not be further discussed in
the paper.

\section{\label{sec:Technical-Design}Technical Design and Implementation}

\subsection{Hardware}

\noindent \textbf{\emph{Overall Design.}} VH tags are based on the
low-power Texas Instruments CC1310 or CC1350 radio-frequency microcontroller
(RF MCU; CC13X0 for short) chip. The chips contain an ARM Cortex-M3
processor, a VHF/UHF data transceiver, an ARM Cortex-M0 processor
dedicated to the radio stack (not user programmable), a low-power
processor called a \emph{sensor controller} that can eliminate relatively
slow (and hence power hungry) wakeups of the M3 processor, and a range
of peripherals, including timers, UART, I2C, and SPI. The chips come
in 4-by-4, 5-by-5, and 7-by-7~mm packages; our tags use the 4-by-4~mm
package.

\begin{figure}
\includegraphics[height=3cm]{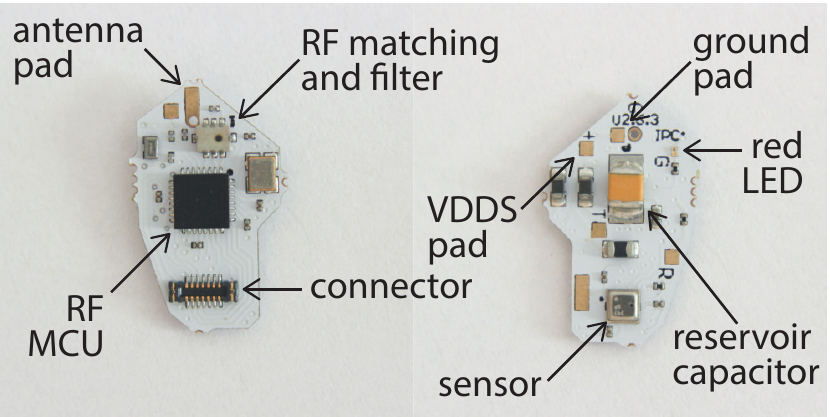}

\caption{\label{fig:tag-anatomy}The main components of a tag (version 2.6.3).}

\end{figure}

All tags contain a number of additional components, including an RF
matching network, to match the RF MCU to the antenna and to filter
harmonics, decoupling capacitors and passives for the switching and
linear regulators, a fairly large (330$\,\mu\text{F}$) \emph{reservoir
capacitor}, a 32~kHz crystal for the real-time clock, a 24~MHz crystal
for the radio, and a miniature male 14-pin board-to-board connector
from the Molex \emph{SlimStack} series. This connector is used for
programming and configuring tags, as well as for attachment of add-on
boards. Figure~\ref{fig:tag-anatomy} shows the parts of a typical
tag.

\begin{figure}
\includegraphics[width=1\columnwidth]{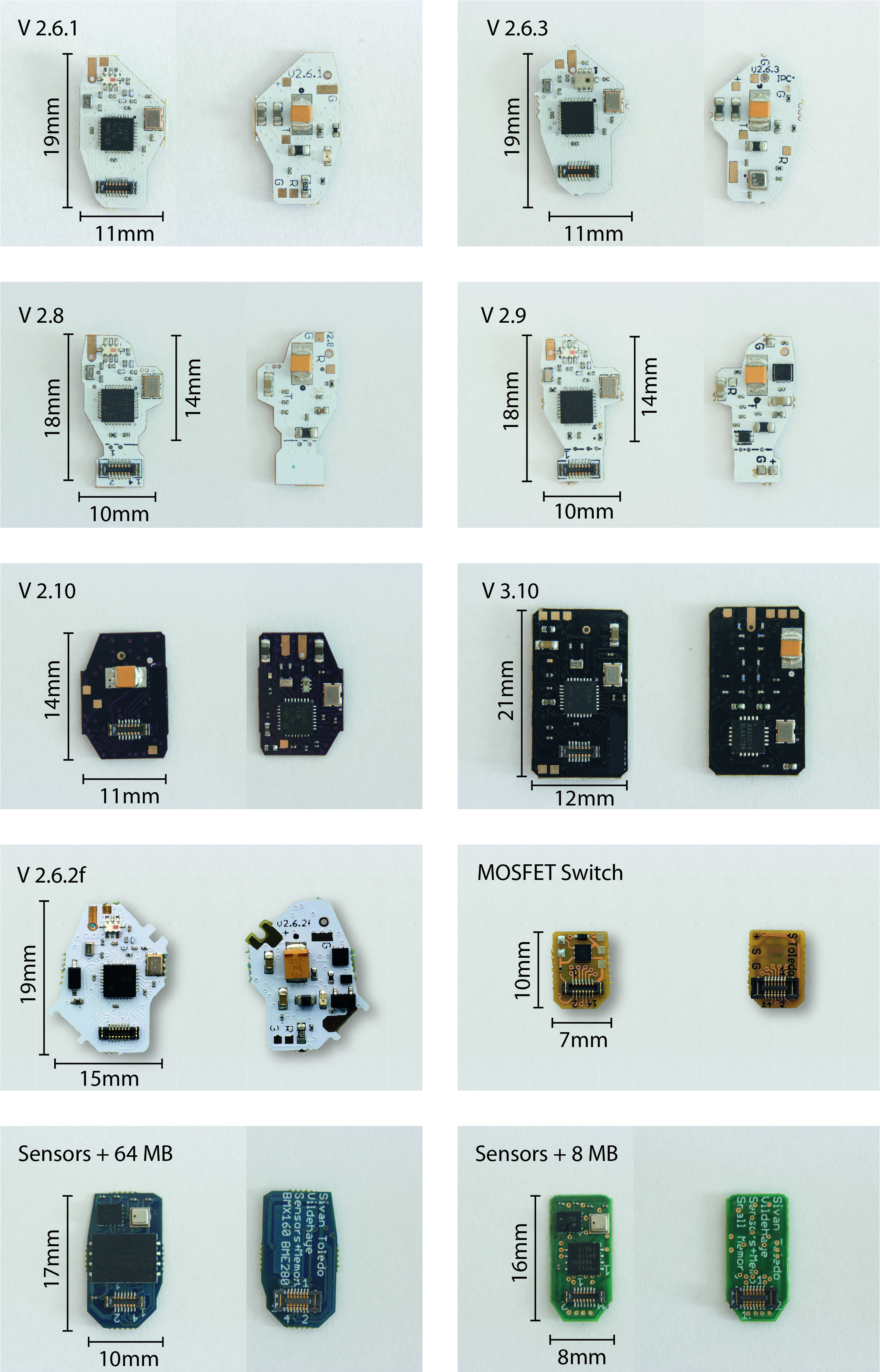}

\caption{\label{fig:tag-catalog}Vildehaye CC13X0 tags and add-on boards. }
\end{figure}

\noindent \textbf{\emph{Hardware Variants.}}\textbf{ }Some tag designs
include additional components, as shown in Figure~\ref{fig:tag-catalog}.
Most versions include an LED that helps users verify that tags function
correctly. Version~2.6.3 includes an on-board air-pressure, temperature,
and humidity sensor (BME280). Versions~2.8 and~2.9 include perforations
between the connector and the rest of the tag, allowing the connector
to be snapped off after programming, to save weight. Their 14-pin
connector does not carry the I2C and SPI buses, to make the boards
more compact. Version~3.10 includes a separate transmitter, AX5043,
to support phase modulation~\citep{modulation}. 

The RF matching networks of the most tag designs are tuned for the
434~MHz band, but version~2.10 is tuned to the 868 and 915~MHz
bands. Some variants (2.6.1, 2.6.2f, 2.8, 2.9) use a discrete balanced-to-unbalanced
network (BALUN) and an integrated low-pass filter, but we also have
all-discrete designs and designs (2.6.3, 3.10) that rely on a single
integrated passive component (IPC) for all the matching and filtering
(2.6.3, 2.10). Most  of our tag designs are shown in Figure~\ref{fig:tag-catalog},
along with some of our add-on boards.

The firmware for both tags and base stations runs on CC13X0 LaunchPads
and on newer CC13X2 LaunchPads, in which the MCU has a stronger processor
(Cortex-M4F) and optionally a 20~dBm RF power amplifier, as well
as on low-cost CC1310 and CC1352 modules from a Ebyte, a Chinese manufacturer.

\noindent \textbf{\emph{Batteries and Voltage Regulation}}. VH tags
are powered by primary batteries including Lithium Manganese Dioxide
coin cells~\citep{Energizer:CR1025,Energizer:CR1620,Energizer:CR2032,Panasonic:CR2477},
pairs of Silver Oxide cells~\citep{Energizer:317,Energizer:337},
and Lithium Thionyl Chloride batteries~\citep{Tadiran:TL4920}. Zinc
Air batteries, which appear effective in laboratory testing~\citep{ToledoIETWSS2015}
proved so far unreliable in the field. All of these batteries can
directly power CC13XX chips, which require a supply voltage and I/O
voltage of 1.8-3.8~V (the Lithium Thionyl Chloride batteries must
be drained a bit before connecting to a tag, to bring their voltage
down to 3.8~V). The chips regulate the supply voltage down to 1.65~V
or 1.8~V (allowing transmission at 10 or 14~dBm, respectively) using
either a linear or a switching regulator. We use the more power-efficient
switching regulator. Most of the internal functional blocks require
even lower voltages generated by internal linear regulators from the
1.65 or 1.8~V rail.

Miniature batteries cannot provide the instantaneous current required
during transmission; this current must be supplied by reservoir capacitors~\citep{BATSReservoirCapacitor,TagsEDERC2014,ToledoIETWSS2015}.
We usually use a 330$\,\mu\text{F}$ tantalum capacitor. Physically
small high-capacitance capacitors are leaky~\citep[F95 Series]{AVX:PolyTantNio};
their leakage current far exceeds the approximately 1$\,\mu\text{A}$
that the MCU consumes in sleep mode. This has two implications: (1)
very low duty cycles are ineffective, because they waste most of the
battery's energy on leakage in the reservoir capacitor, and (2) if
the battery is small, either the capacitor or the battery must be
disconnected until the tag is about to be deployed; we refer to this
as shipping mode.

\noindent \textbf{\emph{Shipping Mode.}}\emph{ }We support three different
shipping-mode mechanisms. The simplest one involves leaving part of
one of the wires connecting the battery to the PCB outside the tag's
coating. The tag is tested and then the wire is cut. To deploy, the
wire is soldered back and the joint is coated with a bit of varnish
or epoxy. This can be done in the field immediately prior to deployment.
The wire is used as a primitive and lightweight switch. This simple
method has two drawbacks. Soldering in the field and covering the
joint are skills that not all ecologists have. Also, with very small
batteries the tag might not start up properly, because the reservoir
capacitor slows down the voltage rise that the MCU senses; the reset
mechanism is not always triggered.

\begin{figure}
\begin{centering}
\includegraphics[clip,scale=0.65]{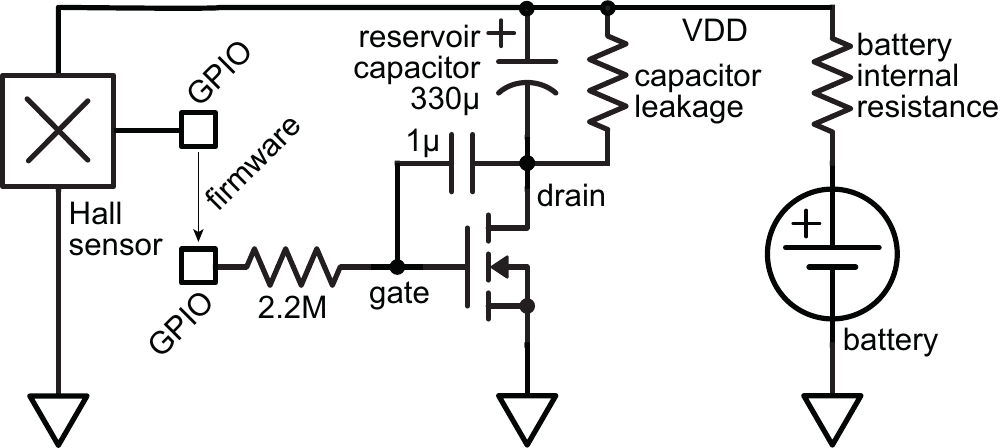}\medskip{}
\par\end{centering}
\begin{centering}
\includegraphics[width=1\columnwidth]{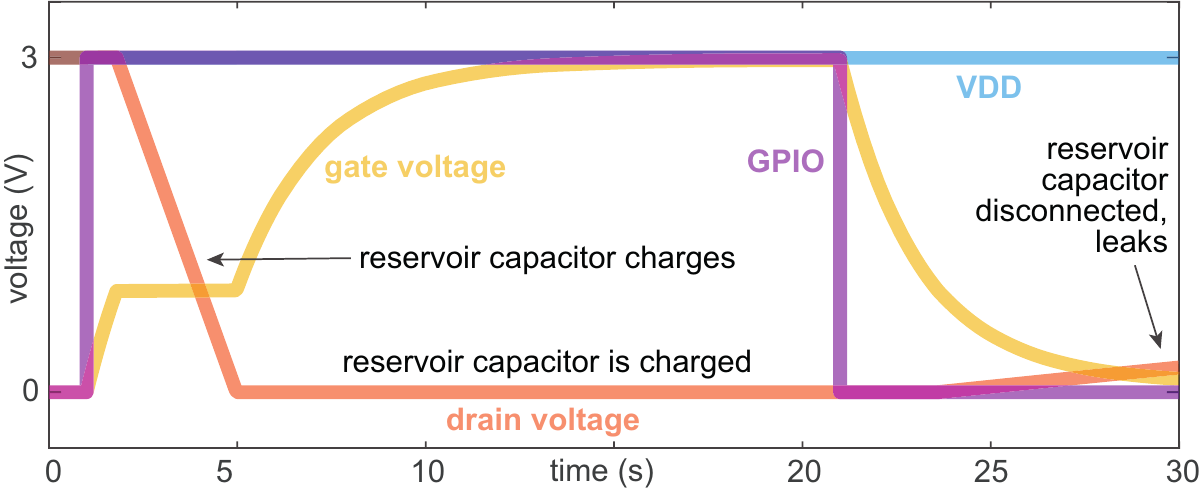}
\par\end{centering}
\caption{\label{fig:Reservoir-Capacitor-Switch}Slowing down the charging of
the reservoir capacitor. The leakage resistor and battery's internal
resistor are not actual components but rather part of the model used
to simulate the behavior (bottom). The simulation used values of 250K
and 10 for the two resistors and was carried out in LTSpice~XVII;
the results were validated experimentally.}
\end{figure}
The second shipping-mode mechanism, shown in Figure~\ref{fig:Reservoir-Capacitor-Switch},
is implemented in Version 2.9. It includes a Hall sensor and a MOSFET
that can disconnect the leaky reservoir capacitor. When a magnet is
placed near the tag, the Hall sensor senses it and the MCU turns off
the MOSFET to disconnect the capacitor. When the magnet is removed,
the reservoir capacitor is reconnected, allowing the tag to transmit.
 Turning on the MOSFET quickly would reconnect the discharged capacitor,
a very low-impedance load, to the miniature battery, causing a sharp
voltage drop that would turn off the MCU. The addition of a large
gate resistor and a gate-drain capacitor amplifies the Miller effect
and charges the capacitor at a controlled constant current, eliminating
the problem.

\begin{figure}
\begin{centering}
\includegraphics[clip,scale=0.65]{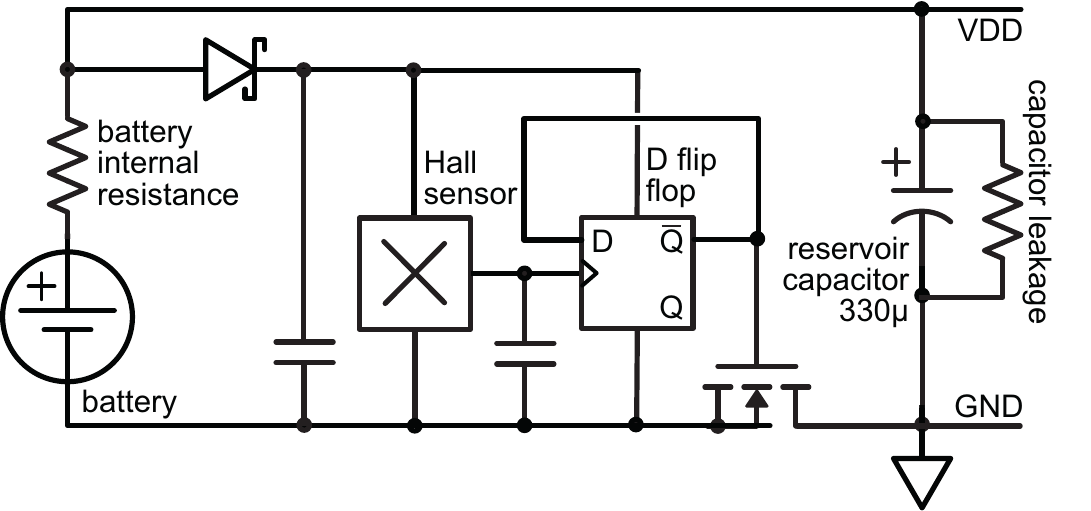}
\par\end{centering}
\caption{\label{fig:Hall-switch}Disconnecting and reconnecting the battery
to the MCU and reservoir capacitor. No firmware support is required.}
\end{figure}
The third shipping mode mechanism, designed for larger batteries (e.g.,
CR2032) and shown in Figure~\ref{fig:Hall-switch} is implemented
in Version 2.6.2f. Here the hall sensor drives a flip flop whose output
connects or disconnects the battery from the rest of the tag, including
both the MCU and the reservoir capacitor. No firmware support is needed,
but the circuit charges the capacitor quickly and it requires an extra
component, the flip flop. 

\noindent \textbf{\emph{Add-on Boards.}} We currently have two types
of add-on boards. One contains an SPI flash memory chip and two sensors,
an inertial measurement unit (IMU) containing an accelerometer, gyroscope,
and optionally a magnetometer (Bosch BMI160 or BMX160), and an air
pressure, temperature, and humidity sensor (Bosch BME280). The memory
chip is either a 64~MB NOR flash in an 8-by-6~mm 8-WSON package
or an 8MB flash in a 4-by-4~mm USON package. All the chips in these
add-on boards allow for a wide range of supply voltages, at least
1.8-3.6~V, almost the same as the RF MCU. The larger 8-WSON footprint
can also accommodate NAND flash chips with much larger capacity, but
these require voltage regulation and firmware features that we have
not yet implemented. 

The other type of add-on board includes a MOSFET low-side switch and
a flyback diode and is designed to activate a DC motor that powers
a tag-release mechanism. 

Add-on boards have a male 14-pin connector on the component side and
a female connector on the back, where no components are placed, to
allow stacking.

Section~\ref{sec:Lessons-Learned} discusses the tradeoffs involved
in the use of add-on boards versus specialized integrated tag variants.

\noindent \textbf{\emph{Manufacturing.}} Tags and other boards in
the VH system have been designed by the authors and have been manufactured
by CircuitHub~\citep{www:circuithub}. We usually use thin 0.4~mm
boards to reduce weight, with 4 copper layers for tags and 2 for add-ons.
Complete tag PCBs cost about 25~USD in batches of 100 and about 11~USD
in batches of 1000 (exact prices vary with the specific design and
over time). Table~\ref{tab:circuithub-runs} shows how many orders
(manufacturing runs) were placed for different VH boards.

\begin{table}
\caption{\label{tab:circuithub-runs}The number of CircuitHub manufacturing
runs for the main VH circuit boards. }

\centering{}%
\begin{tabular}{>{\centering}p{0.5cm}>{\centering}p{0.5cm}>{\centering}p{0.5cm}>{\centering}p{0.5cm}>{\centering}p{0.5cm}>{\centering}p{0.5cm}>{\centering}p{0.5cm}>{\centering}p{0.5cm}>{\centering}p{0.5cm}>{\centering}p{0.5cm}}
\begin{turn}{45}
2.0
\end{turn} & \begin{turn}{45}
2.6.1
\end{turn} & \begin{turn}{45}
2.6.3
\end{turn} & \begin{turn}{45}
2.8
\end{turn} & \begin{turn}{45}
2.9
\end{turn} & \begin{turn}{45}
2.10
\end{turn} & \begin{turn}{45}
3.10
\end{turn} & \begin{turn}{45}
\begin{varwidth}[t]{\linewidth}
\setlength{\baselineskip}{0.5\baselineskip}sensors\\
+64MB
\end{varwidth}
\end{turn} & \begin{turn}{45}
\begin{varwidth}[t]{\linewidth}
\setlength{\baselineskip}{0.5\baselineskip}sensors\\
+8MB
\end{varwidth}
\end{turn} & \begin{turn}{45}
adapter
\end{turn}\tabularnewline
\hline 
1 & 13 & 2 & 7 & 7 & 1 & 1 & 4 & 1 & 4\tabularnewline
\end{tabular}
\end{table}

\noindent \textbf{\emph{Programming.}} Tags are programmed and configured
through a UART bootloader present in all CC13XX RF MCUs~\citep{TI:swra466d}.
For programming, the tag is attached to an FTDI USB-to-UART bridge
using a simple adapter board with a female 14-pin connector. The bridge
also provide two GPIO pins that our Windows-side flashing software
uses to reset the MCU and to activate the bootloader. The bridge is
inexpensive (about 10~USD) and has excellent drivers, simplifying
the task of programming blank MCUs for users; no specialized JTAG
equipment is required. The 14-pin connector does not carry debugging
(JTAG) signals. To debug the firmware, we use Texas Instruments evaluation
boards called \emph{LaunchPads}.

\noindent \textbf{\emph{Base Stasions}}. VH base stations usually
use a LaunchPad evaluation board, sometimes attached to additional
hardware modules. \emph{Tethered }base stations that attached to a
PC or a Raspberry Pi typically contain no hardware beyond the LaunchPad,
which includes a UART-to-USB bridge. Standalone logging base stations
also includes an SD card socket for the log, a u-blox GNSS module
(mainly to set the time), an I2C OLED display, and optionally a temperature
and air-pressure sensor. We have a custom board with all of these
components, designed to attach to a LaunchPad, but it base stations
can also be assembled from a LaunchPad and hardware modules available
from vendors such as Sparkfun and Adafruit or on Ebay. Standalone
base stations that are used for remote command and control of tags
require nothing beyond the LaunchPad, and can also use a tag PCBs. 

\noindent \textbf{\emph{Shortcomings of the CC13XX RF MCUs}}. The
CC13XX RF MCUs serve us well and are extremely well supported by the
manufacturer, but with a few additional or modified features they
would have served us even better. The most important issue is voltage
regulation. The chips regulate a 1.8--3.8~V supply down to 1.65
or 1.8~V using an efficient step-down switching regulator. However,
the regulator cannot supply other devices, so using a NAND flash chip
(all of which require regulated voltage) requires an additional regulator,
adding complexity and weight. Also, the fact that the regulator cannot
be configured as a step-up regulator prevents us from powering tags
with a single 1.5~V cell. Finally, the chips do not tolerate supply
voltages lower than 1.8~V, even though most of the internal blocks
use much lower voltages. Extending the supply voltage down to 1.4
or 1.2~V or less, even with some functional blocks disabled (e.g.
the radio transmitter), would have made VH tags more reliable (see
next section).

The API that configures the radio on CC13XX is only partially documented;
almost all configurations require passing to the radio processor code
patches and/or arrays of parameters that only the vendor can produce
and which are specific to some set of chips (e.g., CC13X0 but not
CC13X2, etc). This leads to complicated and error-prone radio-configuration
code. 

The chips do not support binary phase-shift keying (BPSK), a form
of modulation that is particularly useful for time-of-arrival measurements~\citep{modulation}.
BPSK is easy to produce. We do not know why it is not supported; the
most likely reason is a misguided believe that it is more difficult
to demodulate than frequency-shift keying (FSK); while it is true
that coherent BPSK demodulation might be too complex for simple RF
MCUs, incoherent demodulation of BPSK is just as easy as demodulation
of FSK~\citep{modulation}. 

\subsection{Firmware}

\textbf{\emph{Operating System}}. VH firmware is written in C on top
of Texas Instruments' TI-RTOS operating system. TI-RTOS is an embedded
multitasking operating system designed for single-application devices.
Multiple tasks or threads are preemptively scheduled according to
fixed priorities. It does not offer memory protection or preemptive
time-sharing. TI-RTOS has excellent power management capabilities
for ultra low-power systems. It comes with a set of drivers for all
the peripherals of the CC13XX family. TI-RTOS now supports the POSIX
API for many subsystems, such as threads and synchronization, but
when we started the project, it did not, so our firmware mostly uses
TI-RTOS's idiosyncratic APIs. Also, the APIs for the drivers are all
essentially idiosyncratic. Therefore, porting our code to another
embedded operation system would be challenging.

\noindent 

\noindent 

\noindent \textbf{\emph{Concentration-Polarization Tolerant Scheduling}}.
VH tags use a highly specialized scheduler that ensures that the packets
are transmitted at precise intervals and that can tolerate a temporary
inability of a battery to deliver power. 

Precise timing of packet transmission is important because it allows
ATLAS base stations to predict when the next packet from each tag
will be received. The prediction allows the base station to perform
computationally-expensive signal-processing to detect the packet and
estimate its arrival time on a slice of RF samples only slightly longer
than the packet itself. The scheduler uses a timer that is part of
the radio peripheral to time transmissions, leading to transmission
times that are within 100$\,\mu\text{s}$ or less of the predicted
time. However, the firmware does not wait on this timer between activity
slot, but rather uses the ultra-low power sensor controller to wake
up the ARM processor in time for the next slot.

\begin{figure}
\includegraphics[width=0.7\columnwidth]{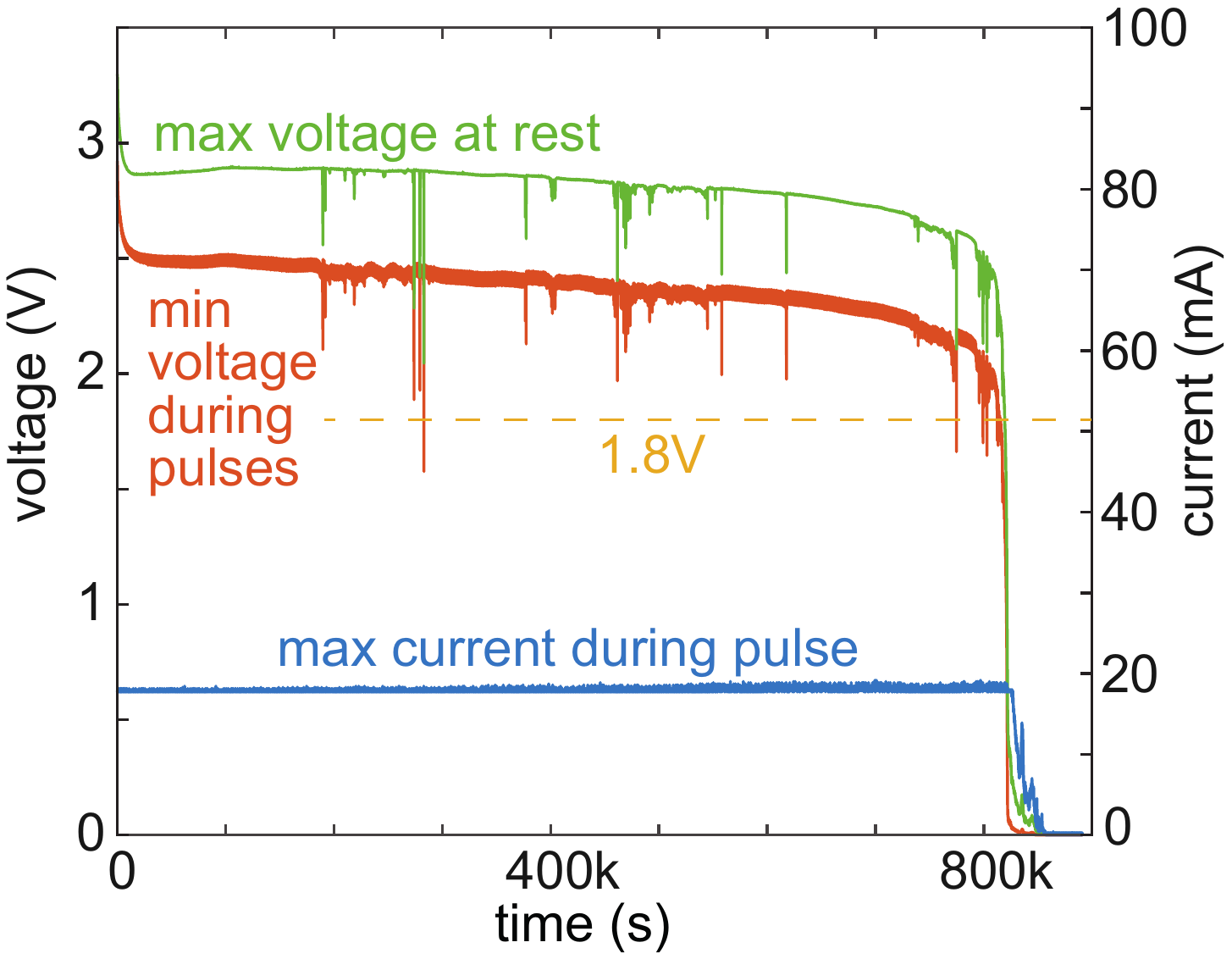}

\caption{\label{fig:Concentration-Polarization}Concentration Polarization
in a Renata CR1025 battery connected to a pulsed load in parallel
with a 330$\,\mu\text{F}$ tantalum capacitor, from~\citep{ShaiMendelMSC}.
Ceramic capacitors make the phenomenon even worse due to lower ESR.}
\end{figure}
The use of the sensor controller allows VH tags to tolerate a phenomenon
called \emph{concentration polarization}, in which internal resistance
increases temporarily because reactants become depleted near the battery's
electrodes (see, e.g.,~\citep{MartinPhD1999,PowerStreamIR}). This
is resolved through diffusion of the reactants, hence resolution can
be slow. This phenomenon can be caused by exposing the battery to
a load impedance load, even for short periods\citep{Peled:private-communication}.
The reservoir capacitor has a low equivalent series resistance (ESR),
so it indeed present a very low-impedance load to the batteries. As
the capacitor is discharged during an activity period, the voltage
across the capacitor drops below the battery voltage, presenting a
very low impedance load to the battery. The graphs in Figure~\ref{fig:Concentration-Polarization},
from~\citep{ShaiMendelMSC}, provide evidence for this phenomenon.
The graphs were produced using a battery load simulator that connected
a 20~mA constant current sink to a Renata CR1025 for 8~ms every
second; this synthesis load is similar to the load presented by a
VH tag. The data shows that the battery voltage sometimes drops significantly
well below the battery is depleted, once below the minimum 1.8~V
required for the CC13XX. Whether this happens vary from battery to
battery, even from the same manufacturing batch~\citep{ShaiMendelMSC}.

VH tags mitigate the risk of failure due to concentration polarization
in two ways. The first is a resistor present on most tag variants
allowing the user to connect the battery through a current-limiting
resistor; the battery never sees a low impedance load, reducing the
risk of concentration polarization. See~\citep{ShaiMendelMSC} on
how to size the resistor. The other is specialized sensor-controller
code that implements the inter activity-period wait. The code sleeps
on a timer. When it wakes up, it measures the supply voltage. If the
voltage has not recovered sufficiently during the wait, the code assumes
that the battery is temporarily unable to provide power, and it waits
until the voltage recovers. Because this firmware mechanism is implemented
using the ultra-low power sensor controller, it has a chance to survive
the concentration polarization phase without causing the battery voltage
to drop below 1.8~V. When the sensor controller finally wakes up
the ARM processor after such a wait, it notifies the ARM processor
that it is off schedule, causing it to restart its scheduler.

\noindent 

\subsection{\label{subsec:sensor-measurements}Representation, Collection, and
Processing of Sensor Measurements}

\textbf{\emph{Representation}}. Tags pack sensor measurements are
packed into \emph{log items}, typed data structures up to 224 byte
long. The length limit allows for efficient on-the-air transport
using the CC13XX radio. The items are stored on nonvolatile memory
and subsequently either uploaded to a base station or read from memory
after the tag is retrieved. The log data structure is described in
detail below.

Data from low-frequency sensors, like a barometric altimeter (one-shot
measurement every second or more), are collected into a log item along
with a 32-bit time  stamp of the first measurement. To save space,
the log item does not store the inter-measurement interval and does
not store the configuration of the sensor (e.g., full scale), because
these are identical in all log items for that sensor. These parameters
are stored in a \emph{sensor-configuration }item on the log every
time the tag boots, and are also available for analysis software from
the file that specifies the tag's definition.

Bursts of samples from high-frequency sensors, like accelerometers
(a 5~s burst at 20~Hz every 2 minutes is typical) do not fit into
a single log item. The samples from the burst are fragmented into
multiple log items. Each item stores a whole-second time stamp of
the beginning of the burst, a one-byte fragment index, and an array
of measurements. Here too, the sampling rate is not represented in
the log items that contain the data from the burst, to save space.

Not storing the sensor configuration in every log item saves on-tag
storage and the energy required to upload the log via radio. Further
saving might be achieved by (lossy or lossless) compression of sensor
measurements; we have not explored this yet.

\noindent \textbf{\emph{Data Collection}}. To process sensor data
from a tag, the log is first transferred to a table in an SQL database.
Each row in the table stores one item in binary form, as well as the
identifier of the tag or base station that generated the item, the
creation time of the log, and the on-flash address of the item in
the log. Software specific to each sensor reads the corresponding
log items from the table, extracts the measurements and associates
each with a time stamp for further processing. 

The log can be transferred to the SQL table in two ways, either by
physically retrieving the tag or via radio upload. When a base station
hears a tag configured to upload data, it invites the tag to upload
data. The base station acknowledges received data items; the tag maintains
a pointer to the last acknowledged log items, to avoid retransmitting
it. The pointer is stored in RAM but is committed to the log in each
sector header. Therefore, a reboot might cause retransmission of log
items, but at most one sector is retransmitted after each reboot.
When a tag leaves the range of a base station, it ceases to upload
data. It may next upload data to another base station into whose range
it enters. Base stations store received log items to an SD card (or
to a file, if the base station is tethered). The cards (or files)
are collected at some point and their contents is uploaded to the
SQL table.

Base stations are fragile and are often placed outdoor, so data that
they collect from tags can sometimes be lost (due to a damaged or
stolen base station, etc). Therefore, the reconstruction of the log
in the SQL table may be incomplete. Software that extracts sensor
data from the table must be and is able to cope with missing items. 

\subsection{A Nonvolatile Log Data Structure}

Tags store sensor data on non-volatile memory using a log data structure.
The log represents a sequence of typed data items up to 224 bytes
long that we refer to as \emph{log items}. The log is stored either
on an external SPI flash chip (currently only NOR flash is supported)
or on the part of the CC13XX flash that is not used for firmware.
The same data structure, stored on raw SD cards (without an underlying
file system) stores packets received from tags in base stations. The
log is currently a write-once data structure that fills an erased
flash chip or SD card; this saves energy when the tag boots. Reusing
flash sectors whose contents have been uploaded to a base station
would consume more energy when a tag boots (see below) but may be
implemented in the future.

\noindent \textbf{\emph{Design Goals}}. The design of the log is optimized
for energy efficiency, storage efficiency, low RAM usage (CC13X0 has
only 20~kB), and clear semantics, including data integrity. Of these
goals, energy efficiency is the most important; it drives most of
the design. 

Note that complete or even high reliability is not one of our design
goals. As explained in Section~\ref{subsec:sensor-measurements},
loss or base stations or thei SD cards may lead to incomplete reconstruction
of the log. Software that uses logged data must be able to cope with
missing data. Therefore, the system can tolerate other causes of loss
of data, as long as they are infrequent. This simplifies the logging
and radio protocols considerably. For example, when a base station
receives a log item from a tag, it normally acknowledges it immediately,
even though writing it to permanent media may fail. However, we obviously
aim to minimize data loss, so a base station with a full SD card does
not acknowledge receipt of log items, and so on.

\textbf{\emph{Naming and Type Labeling}}. Log items, most of which
store sensor data, are globally uniquely identifiable; each is associated
with an abstract 192-bit identifier consisting of the tag identifier,
the UTC time that the physical medium of the log was formatted, and
the address of the item on that medium. The tag identifier and the
log creation time are stored on flash (once for the entire log). The
per-item storage overhead of this unique identifier is close to nothing.
When a log item is uploaded to a base station, the packet containing
it specifies the tag identifier, log creation time, and item address
explicitly. These are also stored in the base stations SD card or
file, and in the SQL table in which data is collected.

The type of a log item is represented by one byte. We currently use
a single registry of log-item types, but the log header contains a
registry identifier. Therefore, even though a single log can contain
only up to 255 different item types, different logs can contain different
sets of types. Most types represent sensor data, but some are part
of the log data structure (e.g., sector headers). 

\noindent \textbf{\emph{Data Integrity}}. Data integrity is achieved
by identifying items that might have been only partially written to
flash due to a power outage. Whenever the tag boots, it logs a \emph{boot
marker}. We ensure that an item has been fully written by verifying
that another item was written later but and before the system lost
power or crashed. Therefore, a data item that is either the last (highest
address) on flash or that is followed by a boot marker is suspect
as being partially written and is not used; other data items have
been written completely and are used by clients. (Enforcing this rule
is more complicated than it seems, since the boot marker may be lost
in transport, but we can identify potential loss of boot markers.).

\noindent 
\begin{figure*}
\begin{centering}
\includegraphics[width=1\textwidth]{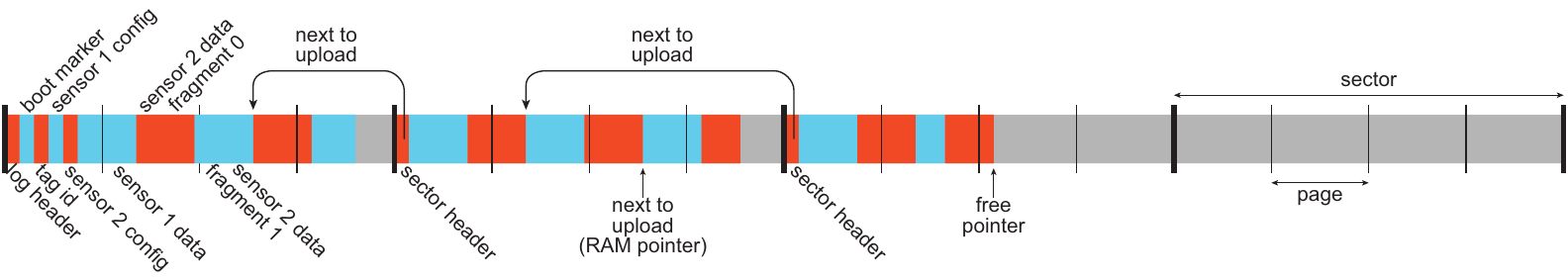}
\par\end{centering}
\caption{\label{fig:log-data-structure}The log data structure, with 4-page
sectors (in practice sectors are larger). The log stores typed \emph{log
items}, shown here as alternating blue and red spans for clarity.
Bytes in the erased state (all ones) are shown in gray. The log starts
with a log-header item. Each sector but the first starts with a sector-header
item. At every boot, the MCU writes a boot marker and its identifier
into the log, as well as items describing the configuration of sensors.
Items are packed contiguously. When the next item to log does not
fit in the space remaining in a sector, writing skips to the next
sector. The address of the first item that has not yet been uploaded
to a base station and acknowledged is maintained in RAM and recorded
in sector headers.}
\end{figure*}
\textbf{\emph{On-Flash Representation of the Log}}. Flash devices
are partitioned into \emph{sectors}, which are the smallest erasable
blocks, which are further partitioned into \emph{pages}. VH logs are
currently not erased, but we still partition the log into logical
sectors (whose size may differ from that of physical sectors). Pages
are 256-byte long on NOR flash and 512-byte long on SD cards. The
implementation never uses byte-write operations, only more energy
efficient page-writes. Each log item is stored with a two-byte header
specifying its length and type. Items are packed without gaps, except
when the next item does not fit within the space left in a sector,
as shown in Figure~\ref{fig:log-data-structure}. 

Because sectors are not erased and rewritten, sectors are filled monotonically
from low to high addresses. This allow code to find the first erased
sector using a binary search, saving energy relative to a linear search
of a large flash chip. The code then scans linearly within the previous
sector to find the first free flash address. 

Write amplification, which costs both storage and energy, is minimal.
Total write amplification consists of the two-byte header (3.2\% for
items 64-byte or longer), sector headers (9 bytes, 0.2\% overhead
for 4~kB sectors), and erased gaps at the end of a sector (at most
225 bytes, 5.5\%). The overhead of the log header and boot marker
are negligible.

Tags upload items to base stations in address order and keep track
of the highest acknowledged address. The tag retransmits the next
item until it is acknowledged. To minimize writes, the tag does not
record each acknowledgment to flash, only when logging operations
start a new sector (when the previous has been filled). 

\subsection{Radio Protocol}

\textbf{\emph{Radio Setups}}. VH tags communicate with base stations
using a simple protocol that uses variable-length radio packets.
The length is encoded in one byte, so the payload is up to 255 bytes
long. The protocol supports a wide range of radio setups, but we normally
use just two: a short-range setup, with symbol rate and data rates
of 500~kb/s and with an error detection code (CRC) but no error correction,
and a long-range setup with the same symbol rate but with spreading
and error-correction codes that reduce the data rate to 31.5~kb/s.
The former is effective in ranges of meters to tens of meters, allowing
energy-efficient data upload to a nearby base station. The latter
was shown to be effective at distances of up to of several kilometers~\citep{vildehaye-wp}
but is only intended for short messages, mostly to identify the tag
and announce its presence in the area.

\noindent \textbf{\emph{Data Representation}}. The payload consists
of a sequence of typed data items, allowing a message to carry multiple
data items. Each data item starts with a variable-length header that
specifies its 16-bit type and length. The header is highly compressed;
if the item is short and the type index low, the header fits in one
byte. Longer data items and high type indexes require longer headers.
Common data types are given low indexes and are kept short, so their
header fits in one byte. 

Packets from tags always carry at least two data items, a \emph{tag
state }data structure and the unique 64-bit tag identifier. The tag-state
structure specifies whether the tag will listen to a reply after this
packet is transmitted, whether it has a significant amount of data
to upload to a base station (the current threshold is 4~KB), and
in what configuration it is. Tags that log sensor data also send every
minute a data item that describes the state of the log, to allow monitoring
the progress of uploads to a base station. Logging tags also send
periodically the time shown by their real-time clock; base stations
reply with the correct time if the advertised time is incorrect (this
is how the clock is set after a tag is powered). Tags in configurations
intended to upload data to a base station, like configuration~1 in
Figure~\ref{fig:tag-def}, include at most one log item in each packet.

\noindent \textbf{\emph{Replies}}. Base stations sometimes reply to
packets from tags. Replies always specify the identifier of the tag
the reply is sent to. Replies are produced by objects called \emph{intents};
as their name suggests, they encapsulate an intent of the user with
respect to a particular tag or to all tags. Every base station that
has a valid real-time clock (either from a GNSS receiver or from NTP)
acts on an intent to adjust the clocks of tags: upon reception of
a packet advertising a local clock that is more than 2~s off from
a tag that will listen next, the base station replies with the correct
time. Similarly, a logging base station that receives a log item acknowledges
it; the tag will move to transmit the next log item. A base station
with a user-specified intent to switch a particular tag to a particular
radio configuration replies to its packet with an appropriate \emph{wakeup
command }data item if the tag's state is not in the intended configuration. 

\noindent \textbf{\emph{Uploading Data}}. Logging base stations that
hear a packet indicating that the tag has data to upload but not carrying
a log item replies with a wakeup command telling the tag to go to
its highest-indexed configuration (without specifying this index,
which the base station does not know). This configuration should be
the one in which data upload occurs (like configuration 1 in Figure~\ref{fig:tag-def}).
This starts a data-upload phase in which the tag sends a log item
almost every period. The base station replies with an acknowledgment
that cause the tag to advance to the next data item. Unacknowledged
items are retransmitted.

\noindent \textbf{\emph{Discussion}}. Many features of the protocol
are designed to save energy and to limit the duration of transmit-receive
slots, to allow the the radio to be powered by a reservoir capacitor.
These include the highly compressed headers and the variable-encoding
of some types of data and the transmission of log items only if a
base station that might acknowledge them was recently heard. Other
features are designed for flexibility. These intents, which make it
relatively easy to add protocol features (e.g., channel or frequency-band
switching), and the structuring of all messages as lists of typed
data items. These goals are sometimes conflicting. For example, a
single fixed message structure would make the protocol more energy
efficient, because no explicit data types would need to be transmitted,
but also much less flexible. We treated flexibility as a constraint
and optimized energy use subject to it.

\noindent \textbf{\emph{A Discarded Design}}. In our initial design
replies did not immediately follow packets from tags but rather occupied
the next activity slot of the tag. The idea was that this would shorten
the time a tag is active in every slot. But this proved difficult
to implement. One difficulty was to time replies precisely, so the
base station would transmit exactly when the tag is receiving. Although
the API of the CC13XX radio driver timestamps received packets and
allows scheduling transmit and receive operations at specific times
in the future, the precise temporal semantics of the API are not well
documented, making implementation of time-slotted protocols. Also,
we did not have a good tracing environment for such protocols; the
environment must be able to show the precise timing of packets using
a separate receiver. Clearly, it is possible to design and implement
time-slotted protocols in sensor networks (see, e.g.,~\citep{TS-LoRa}),
but it requires much more engineering and debugging efforts than immediate
request-reply protocols. The slotted design also requires a complex
radio scheduler for base stations. In our initial implementation a
base station that received a packet and needs to reply in exactly
1~s waits idle during this second. This is inefficient. Scheduling
the reply but performing other radio operations until it is sent is
more efficient but difficult to implement. Therefore, we switched
to the simple and robust protocol described above, in which replies
are almost immediate. 

\begin{table*}[t]
\caption{\label{tab:weights}Weights and lifespans of VH tags with various
batteries and coatings. For battery specifications, see \citep{Energizer:CR1025,Energizer:CR1620,Energizer:CR2032,Energizer:317,Energizer:337,Panasonic:CR2477,Tadiran:TL4920}.}

\centering{}%
\begin{tabular}{llrlrr}
PCB & battery & energy capacity & coating & total mass & max lifespan\tabularnewline
\hline 
2.8 & SO337 (2) & $8\text{.3}\,\text{mAh}\rightarrow2.4\,\text{V}$ & varnish & 0.8~g & 10.4~d @ 1/8~Hz\tabularnewline
 & SO317 (2) & $11.5\,\text{mAh}\rightarrow2.4\,\text{V}$ & varnish & 0.9~g & 13.1~d @ 1/8~Hz\tabularnewline
 & CR1025 & $30\,\text{mAh}\rightarrow2\,\text{V}$ & varnish & 1.2~g & 32~d @ 1/8~Hz\tabularnewline
 & CR1025 & $30\,\text{mAh}\rightarrow2\,\text{V}$ & epoxy & 1.4~g & 32~d @ 1/8~Hz\tabularnewline
2.6 & CR1620 & $81\,\text{mAh}\rightarrow2\,\text{V}$ & epoxy & 2.4~g & 79~d @ 1/8~Hz\tabularnewline
 & CR2032 & $235\,\text{mAh}\rightarrow2\,\text{V}$ & epoxy & 4.2~g & 226~d @ 1/6~Hz\tabularnewline
 & CR2477 & $1000\,\text{mAh}\rightarrow2\,\text{V}$ & epoxy, heat shrink & 10.5~g & 431~d @ 1/8~Hz\tabularnewline
 & TL4920 & $8500\,\text{mAh}\rightarrow2\,\text{V}$ & epoxy \& collar & 90.0~g & \tabularnewline
\hline 
\end{tabular}
\end{table*}

\section{\label{sec:Not-Implemented}Features That Were Considered But Not
Implemented}

Over the coarse of the project we examined but rejected a number of
features that are in general compatible with the project's scope.
This section describes some of these features and the reasons that
we did not implement them. 

\noindent \textbf{\emph{Wakeup Receivers}}. In the summer of 2018
we evaluated an alternative family of RF MCUs, the Flex Gecko family
from Silicon Labs. These MCUs share many characteristics of the CC13X0
from Texas Instruments: they contain an ARM CPU, they can transmit
and receive on the frequency bands that VH tags use, they come in
similarly-sized packages, and they are power efficient. However, they
were advertised as having two additional important features: the ability
to transmit phase-modulated packets and a wakeup receiver. A wakeup
receiver is an ultra-low power circuit that can wake up an MCU when
strong RF signal is detected (some wakeup receivers can detect a particular
pattern, to avoid spurious wakeups, but the wakeup receiver the the
Flex Geckos acts as a tuned power detector). A wakeup receiver allows
low duty-cycle tags to sense the nearby presence of each other without
prior time synchronization~\citep{BATSSensors}.

The evaluation included a test to ensure that the Flex Gecko can transmit
ATLAS pings and a test to evaluate the wakeup receiver. The wakeup
receiver on one evaluation board was able to reliably detect a packet
from another board at up to about 2~m.

However, we did not go on to design VH tags with Flex Geckos due to
several reasons. One was the fact that blank Flex Gecko's do not come
with a serial bootloader; Silicon Labs offers one, but it must be
programmed onto blank chips using JTAG. This would have complicated
the production process for tags. Another important issue was the design
cost involved: using Flex Geckos would require porting the firmware
to a new embedded operating system and designing new tags. Another
damper was the discovery that phase modulation was announced but not
actually supported; we could not even get a target date or any technical
details (data rates in particular are critical for ATLAS localization).

\noindent \textbf{\emph{Battery Tabs, Clips, and Retainers}}. VH tags
are typically powered by a single Lithium cell or by a pair of Silver
Oxide cells. Some Lithium cells come with soldering tabs designed
for soldering the battery to a PCB. One of the tag designs, Version~2.6.2f,
is designed specifically for tabbed CR2032 cells.

However, most Lithium and all Silver Oxide cells lack tabs. Users
connect tabless cells to tags either by soldering thin enamel-coated
wires directly to the cell using a specialized soldering technique
(to avoid heating up the cell, which degrades it), or by spot welding
nickel tabs. These techniques require some expertise and they take
time. 

We evaluated the use of battery clips and retainers, flexible metal
structures that are soldered to a PCB and allow insertion of batteries
of a particular size. We decided not to use clips and retainers, mostly
due to the large variety of batteries that users wanted to use, to
optimize tags for specific animals. To integrate the battery retainer
with the tag, we would have had to design and manufacture a large
variety of tags, one for each battery size. The extra weight of the
retainer and the extra PCB surface area and weight required also contributed
to the decision not to use clips or retainers. Also, some materials
used to weatherproof tags can electrically disconnect the battery
from the clip.

The modular structure of VH tags enables add-on boards with a battery
retainer, but so far there was not demand for them.

\noindent \textbf{\emph{Attaching Batteries using Conductive Glue}}.
In some wildlife tags batteries are attached using conductive glue~\citep{InjectibleUltrasonicTransmitter}.We
experimented with this technique but failed to get it to work well.
Two particular difficulties that we encountered were smears that caused
shorts (a stencil appears to be required) and attachment of the non-PCB
side of the battery, which requires designing a custom metal clip.
These are not insurmountable problems but soldering or spot welding
proved to be good-enough alternatives.

\section{\label{sec:Use-Cases}Use Cases}

\subsection{ATLAS Pingers (Tracking Tags)}

\begin{figure}
\includegraphics[clip,width=0.47\columnwidth]{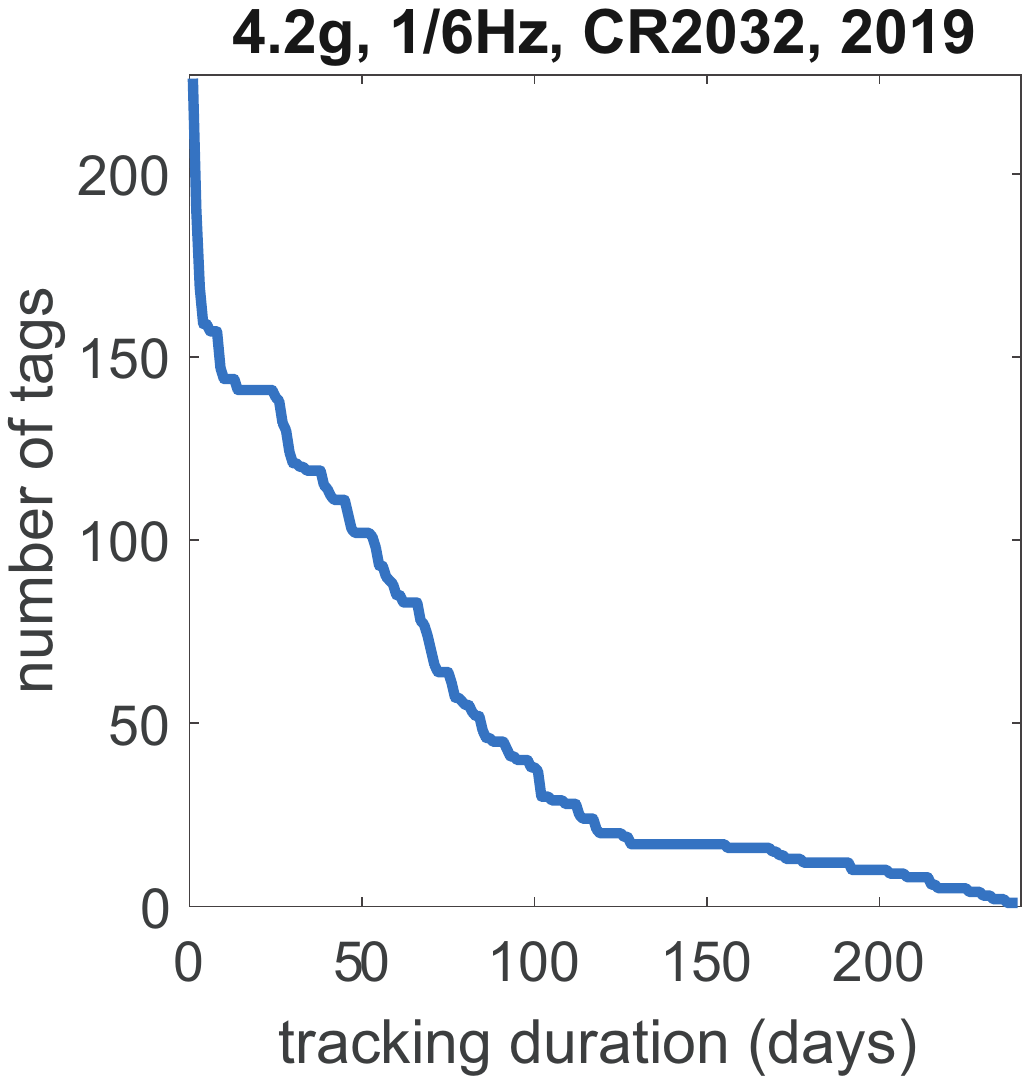}\hfill{}\includegraphics[clip,width=0.47\columnwidth]{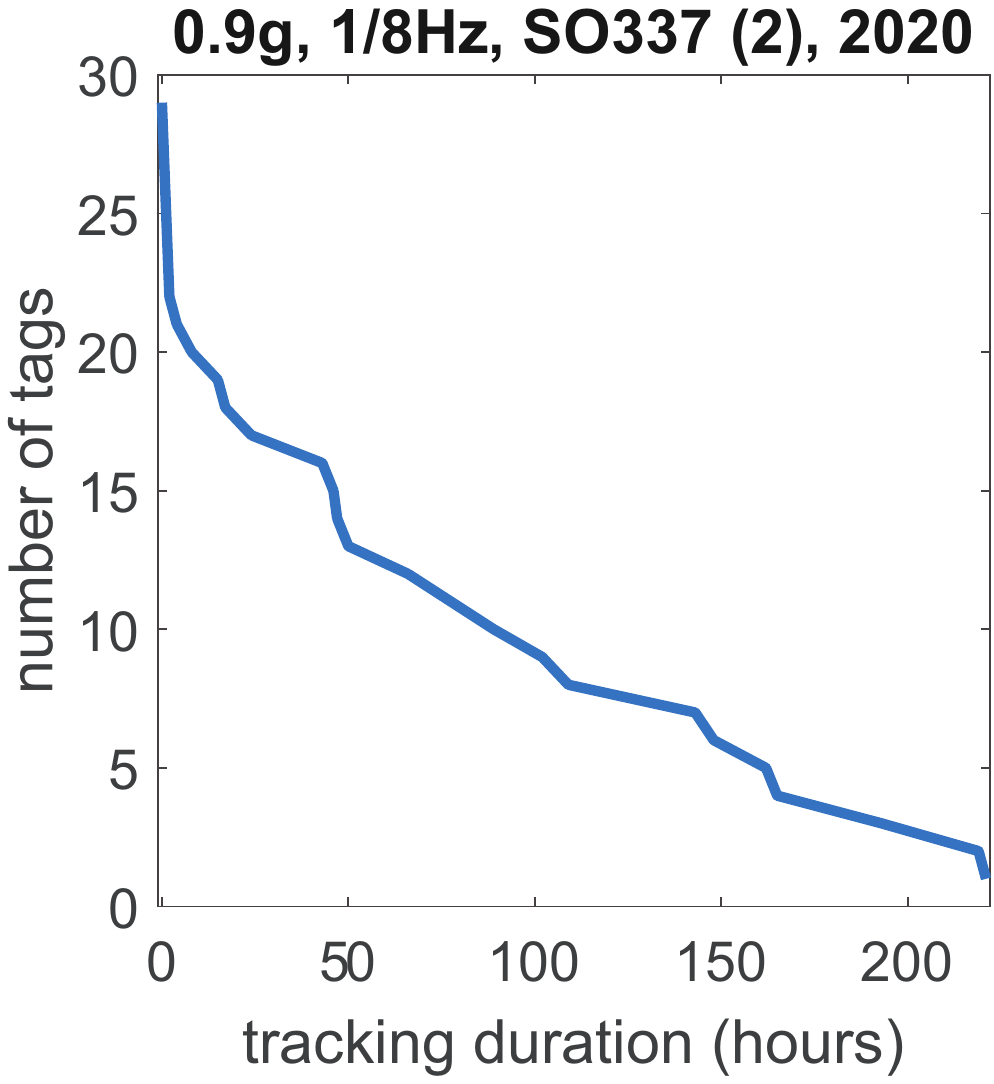}

\caption{\label{fig:tag-life-spans}Tag life spans in the field. The graph
on the left plots results from 226 tags deployed on Red Knots in the
Netherlands and the graph on the right plots results from 29 tags
deployed on Barn Swallows and Common House Martins in Germany.}

\end{figure}
Most of the VH tags that have been deployed in the field have been
configured as ATLAS pingers. Each such tag periodically emits a unique
pseudo-random packet. The packets are transmitted at a high data rate
(usually 8192 bits at 1~Mb/s), to enable accurate localization~\citep{modulation,atlas-accuracy,ChristineValidationArxiv}.
Table~\ref{tab:weights} shows the configuration of typical ATLAS
pingers and the maximum lifespan in the field recorded for each configuration.
Figure~\ref{fig:tag-life-spans} shows the distribution of life spans
among sets of identical tags deployed together. Very early failures
are usually due to defects in tag preparation (e.g. water ingress).
The rest of the distribution is produced mostly by early battery failures
and by animals leaving the coverage area of the ATLAS system (i.e.,
the tag keeps pinging but is not received). It is difficult to ascertain
the cause of the loss of signal from all tags; some user groups collected
such data, but it is incomplete.

\begin{figure*}
\includegraphics[width=0.47\columnwidth]{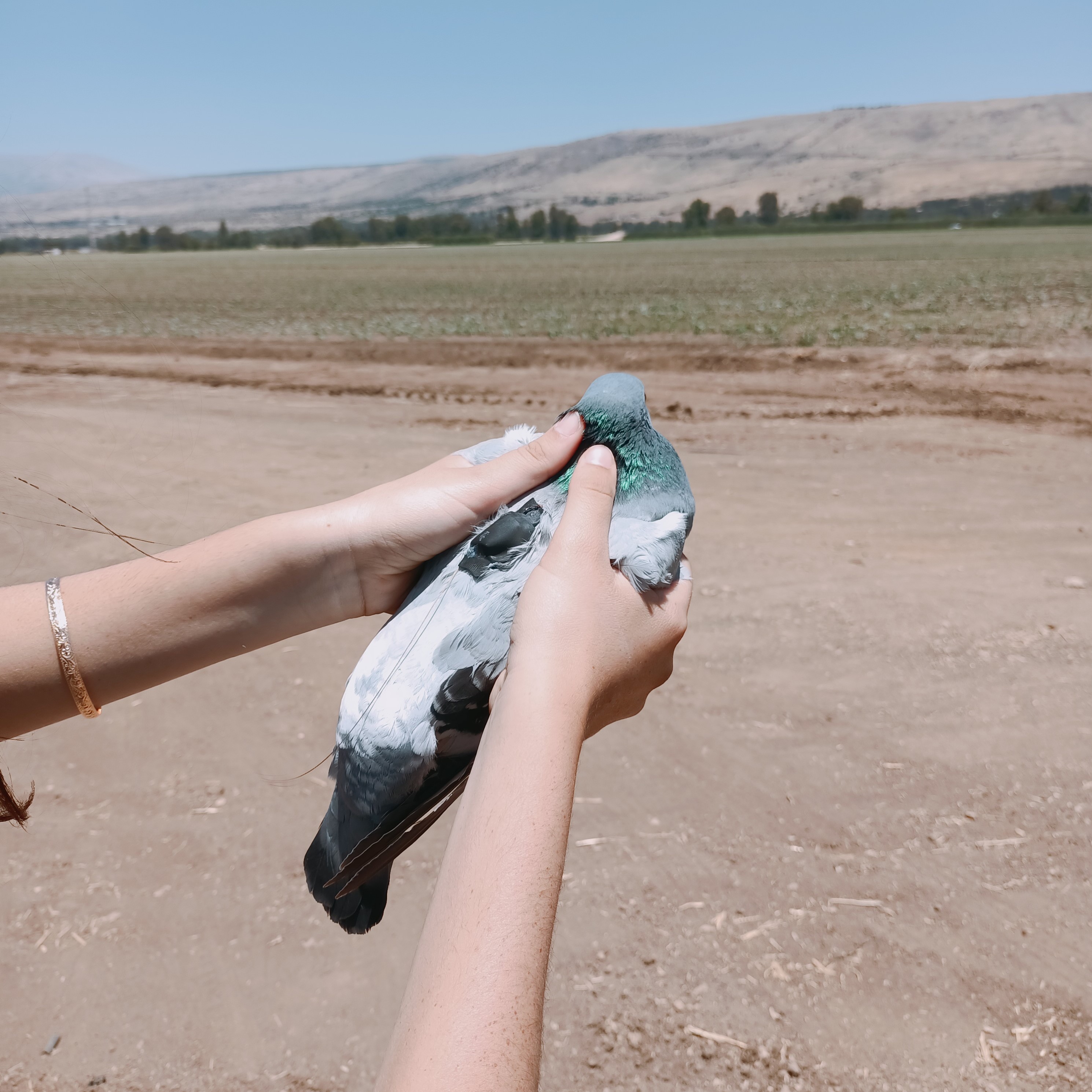}\hfill{}\includegraphics[width=0.47\columnwidth]{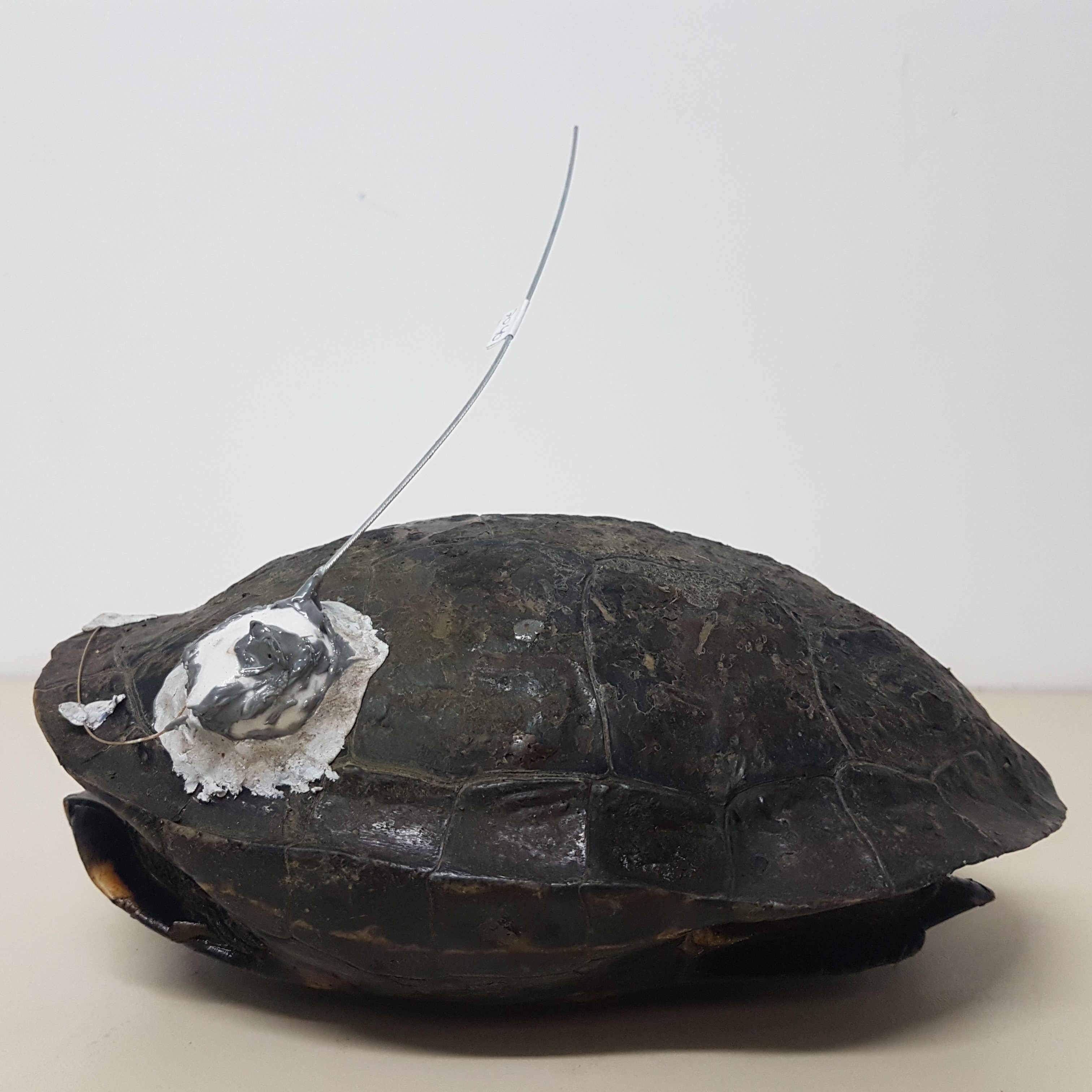}\hfill{}\includegraphics[width=0.47\columnwidth]{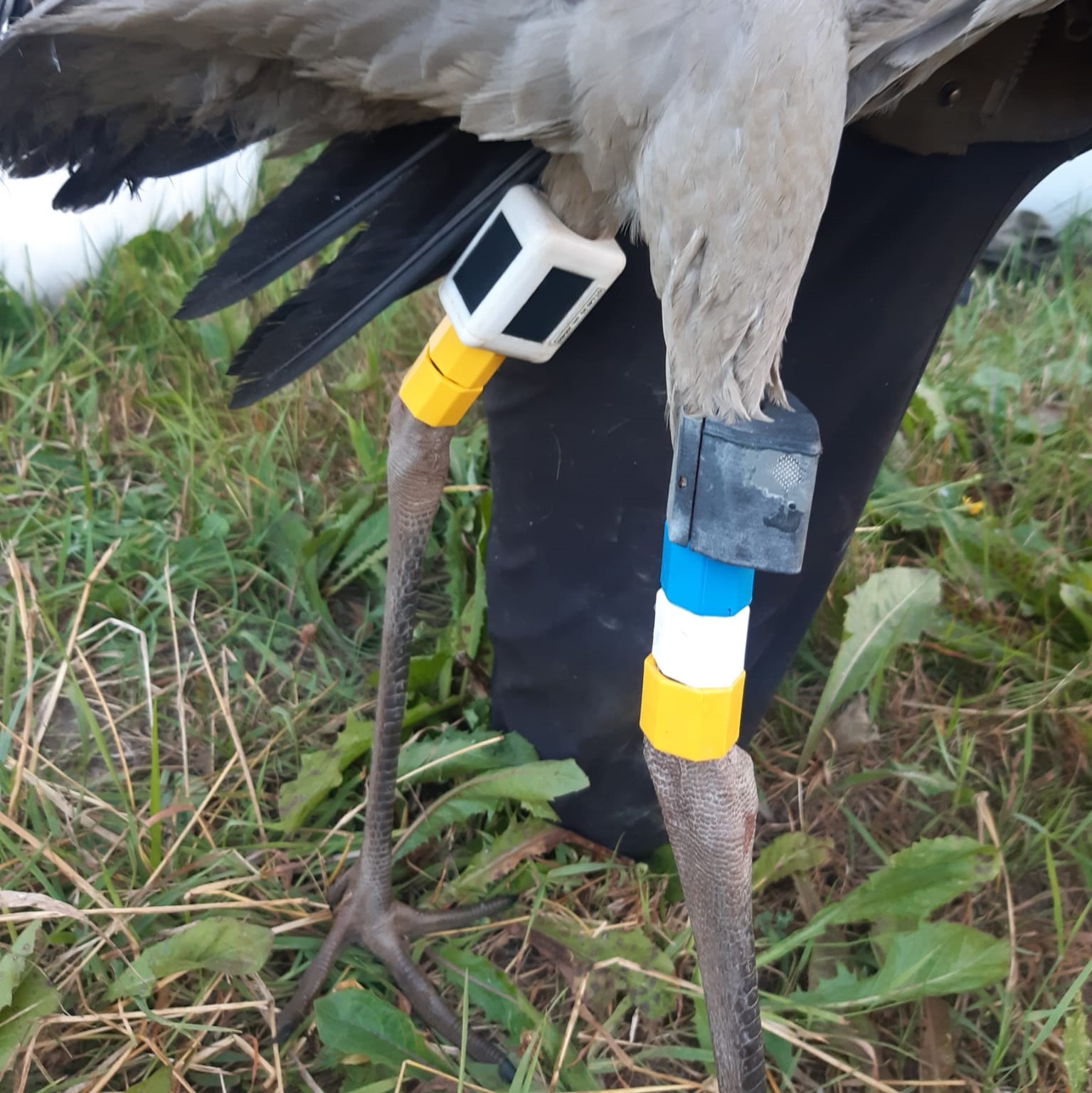}\hfill{}\includegraphics[width=0.47\columnwidth]{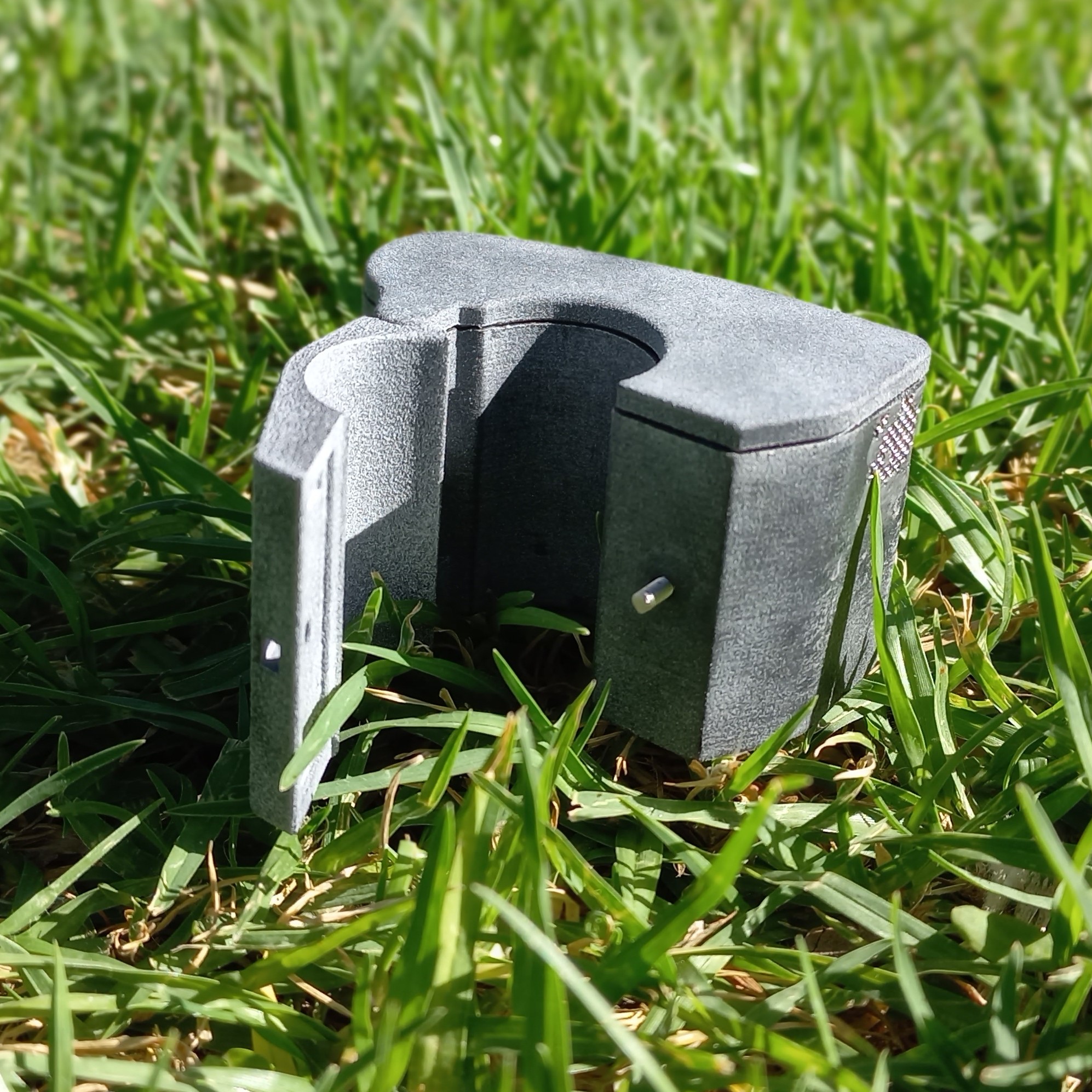}

\caption{\label{fig:tagged animals}A tagged pigeon, terrapin, and crane. The
antenna in the pigeon's tag is a monopole; in the terrapin's tag,
a more effective dipole. The crane carries a data-logging tag with
a remote release mechanism (photo taken by Petri Suorsa), shown in
the released state in the rightmost picture. The VH tag and the data
logger are inside the 3D-printed enclosure. The tag on the other leg
of the crane is a separate GNSS tracker.}
\end{figure*}
Most of these tags have been deployed in ATLAS tracking systems. Figure~\ref{fig:tagged animals}
shows a few tagged animals that were tracked by ATLAS. Most of the
deployed tags were versions 2.6.1, 2.6.2f, 2.6.3., 2.8, and 2.9. 

\subsection{Tracking and Sensing Tags: Recapture and Remote Downloads}

\begin{figure}
\includegraphics[width=0.8\columnwidth]{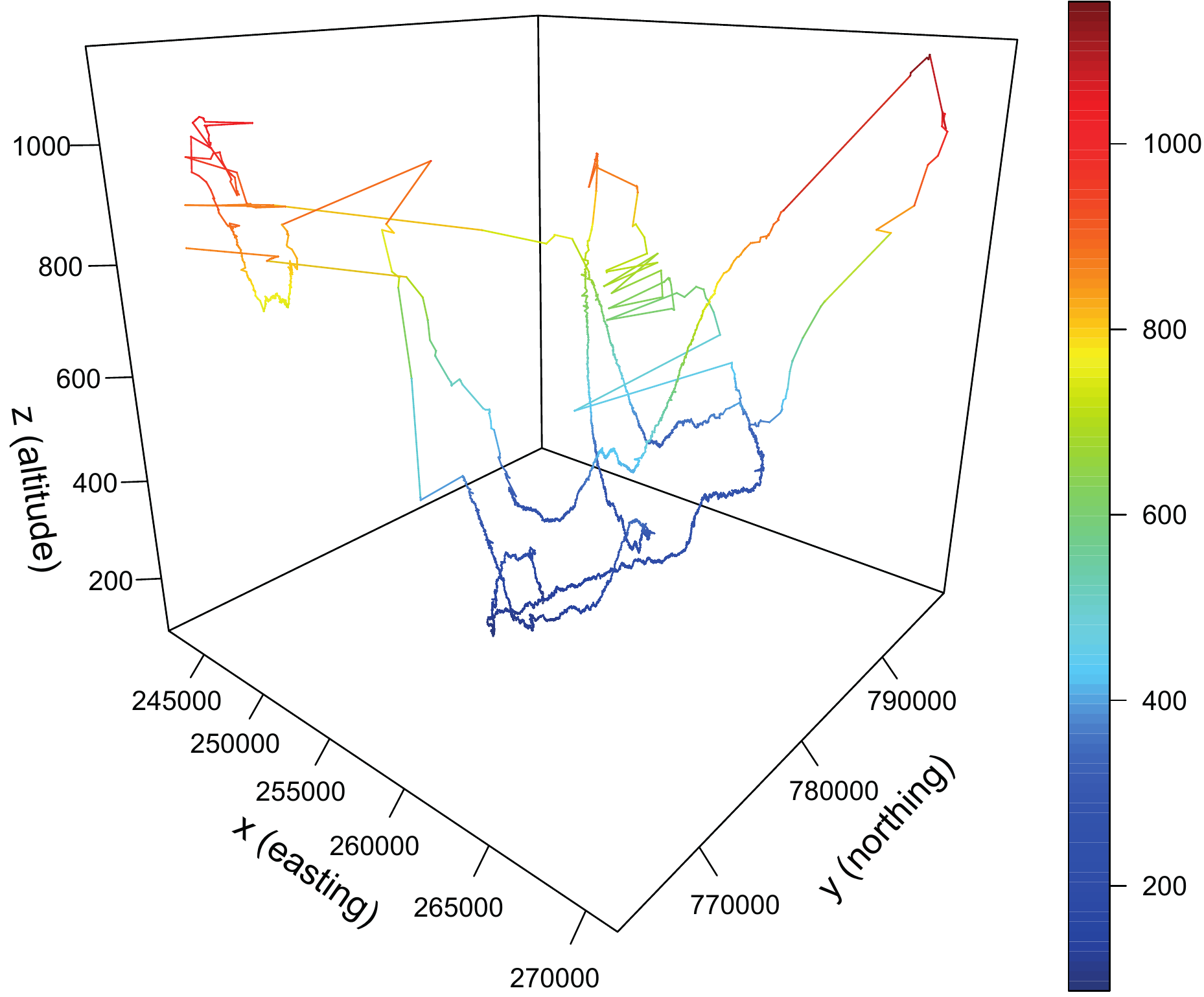}

\caption{\label{fig:pigeon-3d-plot}The 3D track of a pigeon tracked using
a VH tag. Axes labels are in meters. The $x$-$y$ position was estimated
by ATLAS from radio pings and the $z$ coordinate from an air-pressure
sensor whose data was logged on the tag). The $x$-$y$ positions
where not filtered or smooth so the plot shows a few outliers.}
\end{figure}

Two groups of birds were tracked using VH tags that both transmitted
ATLAS pings and logged the air-pressure onto on-board flash memory.
In one study, 27 homing pigeons were tagged with VH tags version 2.6.1
with an add-on board with sensors and a 64~MB flash memory. The birds
were released in remote locations (under multiple experimental conditions
that are irrelevant in this paper) and they flew back home. Upon their
return the tags were removed and the logged data retrieved. Figure~\ref{fig:tagged animals}
shows one of the pigeons just prior to release. Figure~\ref{fig:pigeon-3d-plot}
plots the first part of the track of one of the pigeons in three dimensions.
The air pressure measurements were converted to altitude using the
barometric formula, with reference pressure taken from data published
by a governmental meteorological service. 

Five house martins (\emph{Delichon urbicum}) were tracked in a separate
study  with VH tags version~2.6.3, which have an on-board air-pressure
sensor. The birds nested under the roof of a stable. Two standalone
basestations were placed about 3~m from the nests. Air-pressure data
was remotely uploaded to the base stations and then transferred to
an SQL database for further processing.

We verified the calibration and accuracy of the air-pressure sensor
and the correctness of the estimation of altitude from air pressure
using a drone; we omit the details.

\subsection{Remote Activation of a Release Mechanism}

We used the VH system to design and implement a release-and-retrieval
mechanism for wildlife data loggers. Figure~\ref{fig:tagged animals}
show the mechanism, which we call a \emph{Krukhya}. The data logger
(an audio recorder) and a VH tag are installed in an enclosure with
a motorized flap. The enclosures were placed on the legs a cranes
in Finland and Estonia in the summer with the flap closed, so the
enclosure remains attached to the leg. The cranes were also tracked
using GNSS. When the GNSS tracking indicates that a migrating tagged
crane reached a stopover area in their fall migration, a VH base station
on a drone is flown close to the cranes. It receives long-range transmissions
from the VH tag; it responds with a wakeup command. Receipt of the
wakeup command by the tag turns on the flap motor and also transitions
the tag to a configuration in which it transmits periodic ATLAS pings.
The ATLAS pings are used to home in and locate the released mechanism
using manual RSSI-based direction-of-arrival estimation. 

We have tested the mechanism extensively and determined that a base
station on a drone can activate it from at least 30m (the limiting
factor is the small antenna inside the release-mechanism enclosure).
The actual release in the field is expected to happen later this fall.

\subsection{Remote Download Add-On to Vesper Tags}

We developed a variant of the firmware that assumes that the log is
stored on a separate device accessible as an I2C slave. The VH tag
buffers one log item and tries to upload it to a base station. When
it succeeds, it acknowledges the log item to the logger and asks for
another log item. If there is none, it repeats the request periodically
until the logger produces a log item. The slightly strange setup in
which the logger is an I2C slave rather than a master is due to the
fact that the I2C peripheral on CC13XX devices does not support a
low-power (unclocked) I2C slave or mutiple-master mode.

This mechanism allows the VH tag to act as a remote-upload radio for
existing data loggers. We have integrated this functionality with
a family of wildlife tracking and sensing loggers called \emph{Vesper
}tags~\citep{VesperTags,ASD:Vesper}. This functionality has been
tested and is working, but it was not yet deployed in the field. 

\section{\label{sec:Related-Work}Related Work}

Early wildlife radio tracking tags emitted periodic pings that carried
no information~\citep{CochranLord1963,02g}. Later, simple circuits
that produce a unique on-off pattern were added to allow identification
of individual tags. Tags of this type still attain the lowest mass;
a typical example is the commercial family of NanoTags from Lotek~\citep{www:lotek-nanotags,www:lotek-solar-nanotags},
which start at 0.15~g with either a single Silver Oxide battery or
a solar panel. Such tags are used to either sense the nearby presence
of a tag or to localize it from signal-strength (RSSI) based direction-of-arrival
estimates. 

The next phase in the evolution of radio tracking tags used microcontrollers
to modulate a simple transmitter, either to transmit sensor data or
to identify tags. Lotimer describes an early tag with identifiable
pings~\citep{CodedTransmitterLotimer}; the tag modulated the ping
repetition interval and the number, width, and frequency of pulses
within each ping. The tags developed by MacCurdy et al.~\citep{MacCurdyEtAl2009}
are a more modern, MCU-driven version of this architecture. They were
developed for an automated time-of-arrival localization emit a unique
phase-modulated code. The tags designed by Kr�ger~\citep{Kruger2017}
for a similar system also use the same architecture, but with OOK
modulation.

Integrated transceiver integrated circuits (ICs) enable the design
of yet more sophisticated tags, capable of both transmitting and receiving.
The tags designed for the Encounternet system~\citep{Encounternet}
are typical; they included an MSP430 microcontroller and a CC1101
integrated transceiver. Communication between tags allowed the system
to record short-range encounters between individuals. 

Higher levels of IC integration led to RF MCUs, which allowed for
even smaller tags, including our VH tags. LifeTags and PowerTags~\citep{LifeTagsPLOS,www:ctt-lifetags,www:ctt-powertags}
form a family of periodic 434~MHz pingers that emit a static FSK
packet with the tag ID, designed for presence logging and RSSI-based
direction-of-arrival localization. The tags are single-function and
not configurable. The hardware design appears similar to that of VH
tags, but with an older RF MCU (Silicon Labs Si1060, which has a 8051-compatible
CPU) and a BQ25504 energy harvesting boost converter. One variant
(LifeTags) can operate battery-less using a solar panel and a reservoir
capacitor. The system also include both embedded receivers and hand-held
receivers, like the VH system. The smallest solar LifeTag weighs only
0.45~g, less than VH tags. The functionality is much more restricted
than that of VH tags. The tags are available read-to-use from a commercial
vendor, which is an advantage, but are expensive, costing 180~USD
each at large quantities.

LifeTags, PowerTags, and NanoTags can be detected and identified by
receivers of the Motus system~\citep{Motus}. Motus receivers are
owned and maintained by several research and conservation groups.
Detection reports from all receivers are uploaded to a central database;
the reports can be used to reconstruct an approximate track, sometimes
on continental or intercontinental scales, mostly based on presence
sensing. 

Wildlife tracking using time-of-arrival estimation was tested since
the early 1970s, but early tags where heavy (11.3~kg)~\citep{IRLSBear,IRLSElk}
and the technique was largely abandoned until the work of MacCurdy
et al.~\citep{MacCurdyEtAl2009}.

The BATS project share many similarities with Vildehaye, along with
many contrasting points.~\citep{BATSSensors} BATS is a large-scale
long-term project intended to develop tracking and sensing tags for
bats. VH tags have also been extensively used on bats, but also on
many other animals. BATS tags can be localized using an array of terrestrial
receivers; receivers estimate direction of arrival from RSSI ratios
in two antennas, whereas VH tags are localized by receivers that estimate
time of arrival. BATS tags contain a wakeup receiver, allowing them
to log close encounters between bats. 

\section{\label{sec:Lessons-Learned}Conclusions and Lessons Learned}

Multiple hardware variants and the modular design are a key success
factor because they allow users to select a tag or a modular configuration
that best suits a particular animal and the requirements and constraints
of a particular study. Modular tags with add-on boards have advantages
and disadvantages relative to specialized integrated single-board
variants. Specialized variants require additional design time and
they cost more if they are manufactured in small production runs or
if multiple runs are required to fix faults in a new design (this
has happened to us). On the other hand, integrated variants are physically
smaller and lighter, making them applicable to more species and reducing
adverse effects on tagged animals. For example, version~2.6.2f is
specialized to a particular battery, but a large number has been manufactured,
so the extra costs are minor. Version~2.6.3 is another interesting
example: it is a specialized integrated variant (with an on-board
altimeter), designed to allow tagging small birds that cannot carry
modular tags. 

The decision to manufacture multiple small and medium-size batches
of tags proved effective; we have gone through more than 40 production
runs. This eliminates the need for capital for large runs, enables
using multiple variants, and avoids the risk of a large run of a faulty
design. On the other hand, the multiple small runs expose users and
maintainers to challenges involving part shortages, which we have
faced several times. The reservoir capacitor and 0402 $1\,\mu\text{F}$
capacitors are manufactured by only few vendors and were sometimes
impossible to source. An RF IPC that is specific to the CC13X0 was
impossible to source for a while. These problems were exacerbated
by pandemic-induced shortages (e.g., we can no longer source BME280
sensors and had to switch to BME380, requiring both a board change
and a firmware change), but they also occurred before the pandemic. 

The SlimStack board-to-board connector, adopted from the design of
Vesper tags~\citep{ASD:Vesper}, are effective. The connector is
used both for programming and for communication with add-on boards.
It is tiny, reliable, and easy to use, even in the field. They proved
easier to use than Tag-Connect cables with spring-loaded contacts
that we used to program earlier tags~\citep{TagsEDERC2014}. We were
originally concerned with durability of the connector in the programming
adapter and considered the adapter to be a disposable component, since
the connectors are only rated for 30 mating cycles. However, in practice
we did not experience connector failures. The 1~mm board-to-board
gap requires careful component placement on the PCBs, because in general
only one of the two boards facing a gap can carry components (otherwise
the combined height of components exceeds 1~mm). We place components
on the male-connector side of the board, except on tags Version~2.10
designed to mate with Vesper tags, which have a female connector and
components on the same side. 

The use of two separate serialization formats that Vildehaye uses,
one optimized for data items in radio packets and the other for log
items, is a poor design decision. However, the code is already working
and unifying the two formats will require work without bringing much
benefits.

Planning and executing studies of free-ranging wild animals presents
multiple technical challenges. Every action that we have taken to
simplify the use of VH tags and their associated software contributed
to their adoption and use. These include the (seemingly insignificant)
decision to use an FTDI USB-to-serial bridge to program and configure
tags, the creation and maintenance of the tag lifespan calculator,
and creation of easy ways to test tags after production and before
deployment.

\paragraph*{Acknowledgments.}

This research was supported in part by the Minerva Foundation, the
Minerva Center for Movement Ecology, grants ISF-965/15 and 1919/19
from the Israel Science Foundation, a grant from the Gesellschaft
f�r �kologie, DFG funded research training group \emph{BioMove} (RTG
2118-1), DFG project UL~546/1-1, and Dutch Research Council grant
VI.Veni.192.051. Thanks to the reviewers and shepherd for comments
and suggestions. 

\bibliographystyle{plainurl}
\bibliography{atlas}

\begin{thebibliography}{10}

\bibitem{3M:1601}
3M.
\newblock {\em Electrical Insulating Sealers, 1601-C, 1601-R}, September 2016.
\newblock Datasheet for a clear or red insulating varnish in a spray can.

\bibitem{3M:DP270}
3M.
\newblock {\em Scotch-Weld Epoxy Potting Compound/Adhesive, DP270}, March 2019.
\newblock Datasheet for a clear or black epoxy.

\bibitem{EcologyLettersHabitatSelection}
Christine Beardsworth, Mark Whiteside, Philippa Laker, Ran Nathan, Yotam
  Orchan, Sivan Toledo, Jayden van Horik, and Joah Madden.
\newblock Is habitat selection in the wild shaped by individual-level cognitive
  biases in orientation strategy?
\newblock {\em Ecology Letters}, 24(4):751--760, 2021.
\newblock \href {https://doi.org/10.1111/ele.13694}
  {\path{doi:10.1111/ele.13694}}.

\bibitem{ChristineValidationArxiv}
Christine~E. Beardsworth, Evy Gobbens, Frank van Maarseveen, Bas Denissen, Anne
  Dekinga, Ran Nathan, Sivan Toledo, and Allert~I. Bijleveld.
\newblock Validating a high-throughput tracking system: {ATLAS} as a
  regional-scale alternative to {GPS}, 2021.
\newblock BioRxiv preprint, submitted for publication.
\newblock \href {https://doi.org/10.1101/2021.02.09.430514}
  {\path{doi:10.1101/2021.02.09.430514}}.

\bibitem{SpatialCognitiveAbilityATLAS}
Christine~E. Beardsworth, Mark~A. Whiteside, Lucy~A. Capstick, Philippa~R.
  Laker, Ellis J.~G. Langley, Ran Nathan, Yotam Orchan, Sivan Toledo, Jayden~O.
  van Horik, and Joah~R. Madden.
\newblock Spatial cognitive ability is associated with transitory movement
  speed but not straightness during the early stages of exploration.
\newblock {\em Royal Society Open Science}, 8(3), 2021.
\newblock \href {https://doi.org/10.1098/rsos.201758}
  {\path{doi:10.1098/rsos.201758}}.

\bibitem{meta-analysis-of-biologging-effects}
Thomas~W. Bodey, Ian~R. Cleasby, Fraser Bell, Nicole Parr, Anthony Schultz,
  Stephen~C. Votier, and Stuart Bearhop.
\newblock A phylogenetically controlled meta-analysis of biologging device
  effects on birds: Deleterious effects and a call for more standardized
  reporting of study data.
\newblock {\em Methods in Ecology and Evolution}, 9(4):946--955, 2017.
\newblock \href {https://doi.org/10.1111/2041-210X.12934}
  {\path{doi:10.1111/2041-210X.12934}}.

\bibitem{www:ctt-lifetags}
{Cellular Tracking Technologies}.
\newblock {LifeTag}.
\newblock Retrieved February 9, 2022.
\newblock URL: \url{https://celltracktech.com/products/tag-system/lifetag/}.

\bibitem{www:ctt-powertags}
{Cellular Tracking Technologies}.
\newblock {PowerTag}.
\newblock Retrieved February 9, 2022.
\newblock URL: \url{https://celltracktech.com/products/tag-system/powertag/}.

\bibitem{www:circuithub}
{CircuitHub}.
\newblock Rapid electronics manufacturing.
\newblock Retrieved February 9, 2022.
\newblock URL: \url{https://www.circuithub.com}.

\bibitem{CochranLord1963}
William~W. Cochran and Rexford~T. Lord, Jr.
\newblock A radio-tracking system for wild animals.
\newblock {\em The Journal of Wildlife Management}, 27(1):9--24, 1963.

\bibitem{OwlMicrobial}
Ammon Corl, Motti Charter, Gabe Rozman, Sivan Toledo, Sondra Turjeman,
  Pauline~L. Kamath, Wayne~M. Getz, Ran Nathan, and Rauri C.~K. Bowie.
\newblock Movement ecology and sex are linked to barn owl microbial community
  composition.
\newblock {\em Molecular Ecology}, 20(7):1358--1371, 2020.
\newblock \href {https://doi.org/10.1111/mec.15398}
  {\path{doi:10.1111/mec.15398}}.

\bibitem{IRLSElk}
Frank~.C. Craighead, Jr., John~J. Craighead, Charles~E. Cote, and Helmut~K.
  Buechner.
\newblock Satellite and ground radio tracking of elk.
\newblock In S.~R. Galler, K.~Schmidt-Koenig, G.~J. Jacobs, and R.~E.
  Belleville, editors, {\em Animal Orientation and Navigation}, number 262 in
  NASA Special Publication, pages 99--111. 1972.

\bibitem{IRLSBear}
John~J. Craighead, Frank~C. Craighead, Jr., Joel~R. Varney, and Charles~E.
  Cote.
\newblock Satellite monitoring of black bear.
\newblock {\em BioScience}, 21(24):1206--1212, 1971.
\newblock \href {https://doi.org/10.2307/1296018} {\path{doi:10.2307/1296018}}.

\bibitem{InjectibleUltrasonicTransmitter}
Z.~D. Deng, T.~J. Carlson, H.~Li, J.~Xiao, M.~J. Myjak, J.~Lu, J.~J. Martinez,
  C.~M. Woodley, M.~A. Weiland, and M.~B. Eppard.
\newblock An injectable acoustic transmitter for juvenile salmon.
\newblock {\em Scientific Reports}, 5(8111), 2015.
\newblock \href {https://doi.org/10.1038/srep08111}
  {\path{doi:10.1038/srep08111}}.

\bibitem{BATSReservoirCapacitor}
Falko Dressler, Simon Ripperger, Martin Hierold, Thorsten Nowak, Christopher
  Eibel, Bjorn Cassens, Frieder Mayer, Klaus Meyer-Wegener, and Alexander
  Kolpin.
\newblock From radio telemetry to ultra-low-power sensor networks: tracking
  bats in the wild.
\newblock {\em IEEE Communications Magazine}, 54(1):129--135, 2016.
\newblock \href {https://doi.org/10.1109/MCOM.2016.7378438}
  {\path{doi:10.1109/MCOM.2016.7378438}}.

\bibitem{BATSSensors}
Niklas Duda, Thorsten Nowak, Markus Hartmann, Michael Schadhauser, Bj{\"o}rn
  Cassens, Peter Wägemann, Muhammad Nabeel, Simon Ripperger, Sebastian Herbst,
  Klaus Meyer-Wegener, Frieder Mayer, Falko Dressler, Wolfgang
  Schr{\"o}der-Preikschat, R{\"u}diger Kapitza, J{\"o}rg Robert, J{\"o}rn
  Thielecke, Robert Weigel, and Alexander K{\"o}lpin.
\newblock {BATS}: Adaptive ultra low power sensor network for animal tracking.
\newblock {\em Sensors}, 18(10), 2018.
\newblock \href {https://doi.org/10.3390/s18103343}
  {\path{doi:10.3390/s18103343}}.

\bibitem{VesperTags}
Katya Egert-Berg, Edward~R. Hurme, Stefan Greif, Aya Goldstein, Lee Harten,
  Luis~Gerardo {Herrera M.}, Jos{\'e}~Juan Flores-Mart{\'\i}nez, Andrea~T.
  Vald{\'e}s, Dave~S. Johnston, Ofri Eitan, Ivo Borissov, Jeremy~Ryan Shipley,
  Rodrigo~A. Medellin, Gerald~S. Wilkinson, Holger~R. Goerlitz, and Yossi
  Yovel.
\newblock Resource ephemerality drives social foraging in bats.
\newblock {\em Current Biology}, 28(22):3667--3673, 2018.
\newblock \href {https://doi.org/10.1016/j.cub.2018.09.064}
  {\path{doi:10.1016/j.cub.2018.09.064}}.

\bibitem{Energizer:CR1025}
Energizer.
\newblock {\em Lithium Manganese Dioxide Coin Cell, CR1025}.
\newblock Datasheet, Form No.~1025NA0618; not dated.

\bibitem{Energizer:CR1620}
Energizer.
\newblock {\em Lithium Manganese Dioxide Coin Cell, CR1620}.
\newblock Datasheet, Form No.~1620GL0618; not dated.

\bibitem{Energizer:CR2032}
Energizer.
\newblock {\em Lithium Manganese Dioxide Coin Cell, CR2032}.
\newblock Datasheet, Form No.~CR2032EU0118; not dated.

\bibitem{Energizer:317}
Energizer.
\newblock {\em Silver Oxide Coin Cell, 317}.
\newblock Datasheet, Form No.~317GL0719; not dated.

\bibitem{Energizer:337}
Energizer.
\newblock {\em Silver Oxide Coin Cell, 337}.
\newblock Datasheet, Form No.~337GL0719; not dated.

\bibitem{ATLAS-PRE-PROCESSING}
Pratik~Rajan Gupte, Christine~E. Beardsworth, Orr Spiegel, Emmanuel Lourie,
  Sivan Toledo, Ran Nathan, and Allert~I. Bijleveld.
\newblock A guide to pre-processing high-throughput animal tracking data.
\newblock {\em Journal of Animal Ecology}, 91(2):287--307, 2021.
\newblock \href {https://doi.org/10.1111/1365-2656.13610}
  {\path{doi:10.1111/1365-2656.13610}}.

\bibitem{PheasantCognitionATLAS}
Robert J.~P. Heathcote, Mark~A. Whiteside, Christine Beardsworth, Philippa~R.
  Laker, Jayden Van~Horik, Sivan Toledo, Yotam Orchan, Ran Nathan, and Joah~R.
  Madden.
\newblock Spatial memory and landscape familiarity drive home range formation
  and predict predation risk, March 2021.
\newblock Sumitted for publication.

\bibitem{animal-tags-engineering-perspective}
Mark~D. Holton, Rory~P. Wilson, Jonas Teilmann, and Ursula Siebert.
\newblock Animal tag technology keeps coming of age: an engineering
  perspective.
\newblock {\em Philosophical Transactions of the Royal Society B},
  376(20200229), 2021.
\newblock \href {https://doi.org/10.1098/rstb.2020.0229}
  {\path{doi:10.1098/rstb.2020.0229}}.

\bibitem{Science:GPS-tags}
Roland Kays, Margaret~C. Crofoot, Walter Jetz, and Martin Wikelski.
\newblock Terrestrial animal tracking as an eye on life and planet.
\newblock {\em Science}, 348(6240), 2015.
\newblock \href {https://doi.org/10.1126/science.aaa2478}
  {\path{doi:10.1126/science.aaa2478}}.

\bibitem{Kruger2017}
S.~W. Kr{\"u}ger.
\newblock An inexpensive hyperbolic positioning system for tracking wildlife
  using off-the-shelf hardware.
\newblock Master's thesis, North-West University, South Africa, May 2017.

\bibitem{Kukdo:KH-816}
Kukdo Chemical.
\newblock {\em Curing Agent for use with Epoxy Resins, KH-816}, December 2004.
\newblock Datasheet.

\bibitem{Kukdo:YD-114EF}
Kukdo Chemical.
\newblock {\em Low Viscosity Epoxy Resin, YD-114EF}, December 2004.
\newblock Datasheet.

\bibitem{AVX:PolyTantNio}
Kyocera AVX.
\newblock {\em Polymer, Tantalum and Niobium Oxide Capacitors}.
\newblock Product catalog and datasheets; 285 pages; not dated.

\bibitem{modulation}
Andrey Leshchenko and Sivan Toledo.
\newblock Modulation and signal-processing tradeoffs for reverse-{GPS} wildlife
  localization systems.
\newblock In {\em Proceedings of the European Navigation Conference (ENC)},
  pages 154--165, June 2018.
\newblock \href {https://doi.org/10.1109/EURONAV.2018.8433240}
  {\path{doi:10.1109/EURONAV.2018.8433240}}.

\bibitem{www:lotek-nanotags}
Lotek.
\newblock {NanoTags} (coded vhf) for birds and bats.
\newblock Retrieved February 9, 2022.
\newblock URL: \url{https://www.lotek.com/products/nanotags/}.

\bibitem{www:lotek-solar-nanotags}
Lotek.
\newblock {NanoTags Solar} (coded vhf).
\newblock Retrieved February 9, 2022.
\newblock URL:
  \url{https://www.lotek.com/products/solar-nanotags-coded-vhf-for-birds/}.

\bibitem{CodedTransmitterLotimer}
J.~S. Lotimer.
\newblock A versatile coded wildlife transmitter.
\newblock In Charles~J. Amlaner, Jr. and David~W. Macdonald, editors, {\em A
  Handbook on Biotelemetry and Radio Tracking}, pages 185--191. Pergamon Press,
  1980.

\bibitem{ATLAS-BAT-SPATIAL-PARTITIONING}
Emmanuel Lourie, Ingo Schiffner, Sivan Toledo, and Ran Nathan.
\newblock Memory and conformity, but not competition, explain spatial
  partitioning between two neighboring fruit bat colonies.
\newblock {\em Frontiers in Ecology and Evolution}, 9(732514):1--15, 2021.
\newblock \href {https://doi.org/10.3389/fevo.2021.732514}
  {\path{doi:10.3389/fevo.2021.732514}}.

\bibitem{PowerStreamIR}
Mark~W. Lund.
\newblock Battery impedance and resistance, November 2019.
\newblock URL: \url{https://www.powerstream.com/internal-resistance.htm}.

\bibitem{MacCurdyEtAl2009}
R.~MacCurdy, R.~Gabrielson, E.~Spaulding, A.~Purgue, K.~Cortopassi, and
  K.~Fristrup.
\newblock Automatic animal tracking using matched filters and time difference
  of arrival.
\newblock {\em Journal of Communications}, 4(7):487--495, 2009.

\bibitem{HiddenLivesUndergrandAnimals}
Andrew Markham, Niki Trigoni, Stephen~A. Ellwood, and David~W. Macdonald.
\newblock Revealing the hidden lives of underground animals using
  magneto-inductive tracking.
\newblock In {\em Proceedings of the 8th ACM Conference on Embedded Networked
  Sensor Systems (SenSys)}, pages 281--294, 2010.
\newblock \href {https://doi.org/10.1145/1869983.1870011}
  {\path{doi:10.1145/1869983.1870011}}.

\bibitem{MartinPhD1999}
Thomas~L. Martin.
\newblock {\em Balancing Batteries, Power, and Performance: System Issues in
  {CPU} Speed-Setting for Mobile Computing}.
\newblock PhD thesis, Carnegie Mellon University, 1999.

\bibitem{ShaiMendelMSC}
Shai Mendel.
\newblock A system to characterize battery behavior in miniature wildlife tags
  reveals correctable reliability weaknesses.
\newblock Master's thesis, Tel Aviv University, 2019.

\bibitem{02g}
B.~Naef-Daenzer, D.~Fr{\"u}h, M.~Stalder, P.~Wetli, and E.~Weise.
\newblock Miniaturization (0.2g) and evaluation of attachment techniques of
  telemetry transmitters.
\newblock {\em The Journal of Experimental Biology}, 208:4063--4068, 2005.
\newblock \href {https://doi.org/10.1242/jeb.01870}
  {\path{doi:10.1242/jeb.01870}}.

\bibitem{ScienceReview}
R.~Nathan, C.~T. Monk, R.~Arlinghaus, T.~Adam, J.~Al{\'o}s, M.~Assaf,
  H.~Baktoft, C.~E. Beardsworth, M.~G. Bertram, A.~I. Bijleveld, T.~Brodin,
  J.~L. Brooks, A.~Campos-Candela, S.~J. Cooke, K.~{\O}. Gjelland, P.~R. Gupte,
  R.~Harel, G.~Hellstr{\"{o}}m, F.~Jeltsch, S.~S. Killen, T.~Klefoth,
  R.~Langrock, R.~J. Lennox, E.~Lourie, J.~R. Madden, Y.~Orchan, I.~S. Pauwels,
  M.~{\v{R}}{\'\i}ha, M.~Roeleke, U.~E. Schl{\"{a}}gel, D.~Shohami, J.~Signer,
  S.~Toledo, O.~Vilk, S.~Westrelin, M.~A. Whiteside, , and I.~Jari{\'c}.
\newblock Big-data approaches lead to an increased understanding of the ecology
  of animal movement.
\newblock {\em Science}, 375(6582), 2022.
\newblock \href {https://doi.org/10.1126/science.abg1780}
  {\path{doi:10.1126/science.abg1780}}.

\bibitem{MEE:Magneto-Inductive}
Michael~J. Noonan, Andrew Markham, Chris Newman, Niki Trigoni, Christina~D.
  Buesching, Stephen~A. Ellwood, and David~W. Macdonald.
\newblock A new magneto-inductive tracking technique to uncover subterranean
  activity: what do animals do underground?
\newblock {\em Methods in Ecology and Evolution}, 6(5):510--520, 2015.
\newblock \href {https://doi.org/10.1111/2041-210X.12348}
  {\path{doi:10.1111/2041-210X.12348}}.

\bibitem{Panasonic:CR2477}
Panasonic.
\newblock {\em Lithium Manganese Dioxide Coin Cell, CR2477}.
\newblock Datasheet; not dated.

\bibitem{LifeTagsPLOS}
Teresa~M. Pegan, David~P. Craig, Eric~R. Gulson-Castillo, Richard~M.
  Gabrielson, Wayne~Bezner Kerr, Robert MacCurdy, Steven~P. Powell, and
  David~W. Winkler.
\newblock Solar-powered radio tags reveal patterns of post-fledging site
  visitation in adult and juvenile {Tree} {Swallows} {Tachycineta} bicolor.
\newblock {\em PLoS ONE}, 13(11):e0206258, 2018.
\newblock \href {https://doi.org/10.1371/journal.pone.0206258}
  {\path{doi:10.1371/journal.pone.0206258}}.

\bibitem{Peled:private-communication}
Emanuel Peled.
\newblock Personal communication, October 2018.
\newblock Peled is a Professor of Chemistry in Tel Aviv University.

\bibitem{Encounternet}
Christian Rutz, Zackory~T. Burns, Richard James, Stefanie~M.H. Ismar, John
  Burt, Brian Otis, Jayson Bowen, and James~J.H. {St Clair}.
\newblock Automated mapping of social networks in wild birds.
\newblock {\em Current Biology}, 22(17):R669--R671, 2012.
\newblock \href {https://doi.org/10.1016/j.cub.2012.06.037}
  {\path{doi:10.1016/j.cub.2012.06.037}}.

\bibitem{ASD:Vesper}
Alexander Schwartz.
\newblock {Vesper} loggers family line.
\newblock Retrieved February 9, 2022.
\newblock URL: \url{https://asd-tech.com/products/}.

\bibitem{Tadiran:TL4920}
Tadiran.
\newblock {\em Lithium Thionyl Chloride C-Size Battery, TL-4920}, January 2006.
\newblock Datasheet.

\bibitem{Motus}
Taylor, P.~D., T.~L. Crewe, S.~A. Mackenzie, D.~Lepage, Y.~Aubry, Z.~Crysler,
  G.~Finney, C.~M. Francis, C.~G. Guglielmo, D.~J. Hamilton, R.~L. Holberton,
  P.~H. Loring, G.~W. Mitchell, D.~Norris, J.~Paquet, R.~A. Ronconi,
  J.~Smetzer, P.~A. Smith, L.~J. Welch, and B.~K. Woodworth.
\newblock The {Motus} wildlife tracking system: a collaborative research
  network to enhance the understanding of wildlife movement.
\newblock {\em Avian Conservation and Ecology}, 12(1), 2017.
\newblock \href {https://doi.org/10.5751/ACE-00953-120108}
  {\path{doi:10.5751/ACE-00953-120108}}.

\bibitem{ToledoIETWSS2015}
Sivan Toledo.
\newblock Evaluating batteries for advanced wildlife telemetry tags.
\newblock {\em IET Transactions on Wireless Sensor Systems}, 5:235--242, 2015.
\newblock \href {https://doi.org/10.1049/iet-wss.2014.0042}
  {\path{doi:10.1049/iet-wss.2014.0042}}.

\bibitem{TagsEDERC2014}
Sivan Toledo, Oren Kishon, Yotam Orchan, Yoav Bartan, Nir Sapir, Yoni Vortman,
  and Ran Nathan.
\newblock Lightweight low-cost wildlife tracking tags using integrated
  tranceivers.
\newblock In {\em Proceeings of the 6th Annual European Embedded Design in
  Education and Research Conference (EDERC)}, pages 287--291, Milano, Italy,
  September 2014.
\newblock \href {https://doi.org/10.1109/EDERC.2014.6924406}
  {\path{doi:10.1109/EDERC.2014.6924406}}.

\bibitem{vildehaye-wp}
Sivan Toledo, Yotam Orchan, David Shohami, Motti Charter, and Ran Nathan.
\newblock Physical-layer protocols for lightweight wildlife tags with
  {Internet}-of-things transceivers.
\newblock In {\em Proceedings of the 19th IEEE International Symposium on a
  Wolrd of Wireless, Mobile, and Multimedia Networks ({WOWMOM})}, June 2018.
\newblock \href {https://doi.org/10.1109/WoWMoM.2018.8449778}
  {\path{doi:10.1109/WoWMoM.2018.8449778}}.

\bibitem{Toledo188}
Sivan Toledo, David Shohami, Ingo Schiffner, Emmanuel Lourie, Yotam Orchan,
  Yoav Bartan, and Ran Nathan.
\newblock Cognitive map-based navigation in wild bats revealed by a new
  high-throughput tracking system.
\newblock {\em Science}, 369(6500):188--193, 2020.
\newblock \href {https://doi.org/10.1126/science.aax6904}
  {\path{doi:10.1126/science.aax6904}}.

\bibitem{vilk2021ergodicity}
Ohad Vilk, Yotam Orchan, Motti Charter, Nadav Ganot, Sivan Toledo, Ran Nathan,
  and Michael Assaf.
\newblock Ergodicity breaking and lack of a typical waiting time in
  area-restricted search of avian predators, 2021.
\newblock arXiv preprint, submitted for publication.
\newblock \href {http://arxiv.org/abs/2101.11527} {\path{arXiv:2101.11527}}.

\bibitem{atlas-accuracy}
Adi Weller, Yotam Orchan, Ran Nathan, Motti Charter Anthony~J. Weiss, and Sivan
  Toledo.
\newblock Characterizing the accuracy of a self-synchronized reverse-{GPS}
  wildlife localization system.
\newblock In {\em Proceeings of the 15th ACM/IEEE International Conference on
  Information Processing in Sensor Networks (IPSN)}, Vienna, Austria, April
  2016.
\newblock \href {https://doi.org/10.1109/IPSN.2016.7460662}
  {\path{doi:10.1109/IPSN.2016.7460662}}.

\bibitem{TI:swra466d}
Elin Wollert.
\newblock {\em CC2538/CC26x0/CC26x2 Serial Bootloader Interface}.
\newblock Texas Instruments, August 2021.
\newblock Application Note SWRA466 revision D.

\bibitem{TS-LoRa}
Dimitrios Zorbas, Khaled Abdelfadeel, Panayiotis Kotzanikolaou, and Dirk Pesch.
\newblock {TS-LoRa}: Time-slotted lorawan for the industrial {Internet} of
  things.
\newblock {\em Computer Communications}, 153:1--10, 2020.
\newblock \href {https://doi.org/10.1016/j.comcom.2020.01.056}
  {\path{doi:10.1016/j.comcom.2020.01.056}}.

\end{thebibliography}

\end{document}